\documentclass[aps,superscriptaddress,floatfix, preprintnumbers,nofootinbib]{revtex4}

\pdfoutput=1
\usepackage[T1]{fontenc}
\usepackage{amsmath,amssymb}
\usepackage{epsfig}
\usepackage{subfigure}
\usepackage{float}
\usepackage{graphicx}
\usepackage{bigints}
\usepackage{gensymb}
\usepackage{pifont}
\usepackage[paperwidth=210mm,paperheight=297mm,centering,hmargin=2cm,vmargin=2.5cm]{geometry}
\newcommand{\cmark}{\ding{52}}%
\newcommand{\xmark}{\ding{55}}%
\usepackage[usenames,dvipsnames]{color}
\usepackage{slashed}
\usepackage{multirow}
\usepackage[colorlinks,citecolor=blue]{hyperref}
\usepackage{pdfpages}
\usepackage{color}
\usepackage{comment}
\usepackage{orcidlink}

\newcommand{\mhsqr}[1]{M_{H_1}^2}

\begin{document}
\hfill \preprint {}
\title{Investigating two-zero texture in the light of gauged Type-II seesaw}
\author{Anirban Biswas\,\orcidlink{0000-0002-3810-3326}}
\email{anirban.biswas.sinp@gmail.com}
\affiliation{Department of Physics, Gaya College (A constituent unit of Magadh University, Bodh Gaya),
Gaya, 823001, India}
\author{Shilpa Jangid\,\orcidlink{0000-0001-6307-1234}}
\email{shilpajangid123@gmail.com}
\affiliation{Asia Pacific Center for Theoretical Physics (APCTP)\\
San 31, Hyoja-dong, Nam-gu, Pohang 790-784, Korea}
\affiliation{Shiv Nadar IoE Deemed to be University, Gautam Buddha Nagar,
Uttar Pradesh, 201314, India}
\author{Seong Chan Park\,\orcidlink{0000-0003-0176-4355}} 
\email{sc.park@yonsei.ac.kr}
\affiliation{Department of Physics \& Lab for Dark Universe \&
Institute of Physics and Applied Physics (IPAP),
Yonsei University, 50 Yonsei-ro, Seodaemun-gu, Seoul 03722, South Korea}

\begin{abstract}
Neutrino oscillation, discovered over two decades ago, confirmed that neutrinos have nonzero masses. Since then, two mass-squared differences have been measured with unprecedented precision, yet the absolute neutrino mass scale remains unknown. Additionally, the fundamental symmetry governing the neutrino mixing pattern is still undetermined. Among various theoretical possibilities, the two-zero texture in the neutrino mass matrix ($m_\nu$) stands out as an attractive framework due to its reduced number of free parameters, enabling definite predictions for the unknown parameters of the PMNS matrix. In this work, we present a comprehensive analysis of the two-zero texture, focusing on its implications for the Dirac CP phase ($\delta$), the Majorana phases ($\rho$, $\sigma$) and the effective Majorana mass ($m_{\beta\beta}$), the latter being crucial for neutrinoless double beta decay. We find that for certain two-zero textures, $m_{\beta\beta}$ reaches a few tens of meV, placing it within the sensitivity range of KamLAND-Zen. Furthermore, we demonstrate how a two-zero texture can naturally emerge in a well-motivated neutrino mass model, specifically the gauged Type-II seesaw mechanism, which requires multiple scalar triplets. Notably, some of the two-zero patterns cannot be realized in this framework, as more than two independent zeros appear in $m_\nu$. Finally, we discuss key phenomenological consequences of the gauged Type-II seesaw model.
\end{abstract}

\maketitle
\section{introduction}
\label{sec:intro}
The discovery of neutrino oscillation at the end of the last
century not only solved the atmospheric \cite{Super-Kamiokande:1998kpq, Kajita:2010zz}
and solar neutrino \cite{Nakahata:2022xvq}
problems elegantly but most importantly, it confirmed that neutrinos
have tiny masses with nonzero mass differences among the different
flavours. This was a clear indication of physics beyond the Standard Model (BSM)
since neutrinos are massless in the Standard Model (SM). Till now, more
than two decades after the first discovery of neutrino oscillation,
we precisely know the values of two mixing angles and two
mass square differences \cite{Esteban:2024eli}. The other mixing
angle, called the atmospheric mixing angle
($\theta_{23}$), has a large error bar ($\sim 41^\circ - 50^\circ$ in
$3\sigma$ range). The upcoming long baseline experiments
like DUNE \cite{DUNE:2015lol}, Hyper-Kamiokande \cite{Hyper-Kamiokande:2018ofw}
are expected to resolve this ``Octant degeneracy'' \cite{Agarwalla:2016fkh}
by measuring $\theta_{23}$ more accurately. Moreover, there are other parameters
like the Dirac CP phase\footnote{Generally indicated by
$\delta_{\rm CP}$ in literature.}\,($\delta$), the absolute
mass scale of neutrinos (or equivalently the sum of three
neutrino masses), etc. which we also need to be
understood precisely. On top of that, quest for uncovering
the nature of neutrinos (Dirac or Majorana) is another
biggest issue that may be revealed in experiments
searching for neutrinoless double beta
decay \cite{Dolinski:2019nrj, Jones:2021cga}.
Here is the best fit parameters from the latest
observations\footnote{Note $\Delta{m^2_{3\ell}} = m^2_3 - m^2_1 >0$ for the normal
mass ordering (NO) and $\Delta{m^2_{3\ell}} = m^2_3 - m^2_2 <0$ for the
inverse mass ordering (IO).}
(NuFIT 6.0 (2024))~\cite{Esteban:2024eli}:
\begin{table}[h]
\begin{center}
\caption{Observables in the neutrino sector
(without SK atmospheric data) \cite{Esteban:2024eli}.}
\label{tab:oscillation_data}
\vspace{2mm}\label{oscx}
 \begin{tabular}{|c|c|c|c|c|c|c|}
\hline
\hline
${\rm Parameter}$&$\theta_{12}$&$\theta_{23}$ &$\theta_{13}$ &$
\dfrac{\Delta m_{21}^2}{10^{-5}\,{\rm eV}^2}$&$\dfrac{|\Delta m_{3\ell}^2|}{10^{-3}\,{\rm eV}^2}$
& $\delta$\\
&${(\rm deg)}$&${(\rm deg)}$ &$({\rm deg})$ & & &(${\rm deg})$\\
\hline
$3\sigma\hspace{1mm}{\rm
range\hspace{1mm}(NO)\hspace{1mm}}$&$31.63-35.95$&$41.0-50.5$&$8.18-8.87$&
$6.92-8.05$&$2.463-2.606$ & $96-422$\\
\hline
$3\sigma\hspace{1mm}{\rm
range\hspace{1mm}(IO)\hspace{1mm}}$&$31.63-35.95$&$41.4-50.6$&$8.24-9.21$&
$6.92-8.05$&$2.438-2.584$ & $201-348$\\
\hline
${\rm Best\hspace{1mm}{\rm fit\hspace{1mm}}value\hspace{1mm}(NO)}$ &
$33.68$ & $48.5$ &  $8.52$ &$7.49$ & $2.534$ & 177\\
\hline
${\rm Best\hspace{1mm}{\rm
fit\hspace{1mm}}value\hspace{1mm}(IO)}
$&$33.68$&$48.6$&$8.58$&$7.49$&$2.510$ & 285\\
\hline
\end{tabular}
\end{center}
\end{table}
   
The seesaw mechanism~\cite{Minkowski:1977sc, Yanagida:1979as, Mohapatra:1979ia}
and its variants ~\cite{Schechter:1980gr, Mohapatra:1980yp,
Cheng:1980qt, Foot:1988aq, Ma:1998dx, Arhrib:2011uy, Park:2009cm},
have been suggested to provide the tiny Majorana neutrino masses
by introducing additional heavy species (scalar or fermion) in
the particle spectrum or extra dimensions. The Majorana mass matrix ($m_{\nu}$)
being a complex symmetric matrix has six independent complex
elements, or twelve real elements.

Without knowing the underlying physics of flavor, several approaches have been tried to understand the neutrino masses and their mixings: seesaw mechanisms, radiative mechanisms, flavor symmetries, and extra dimensions. It may not be surprising that some of those approaches are correlated and texture zeros are realized as a consequence of underlying physics. In particular, Two-zero texture \cite{Frampton:2002yf, Xing:2002ta, Fritzsch:2011qv, Ludl:2014axa, Meloni:2014yea, Dev:2014dla} refers to an extremely constrained scenario where one can have two independent zeros in $m_{\nu}$.  This allows fifteen ($^{6}\mathbb{C}_{2}$) different patterns in $m_{\nu}$ out of which only seven are allowed by the neutrino oscillation experiments \cite{Fritzsch:2011qv}.
These are labelled as $A_1$, $A_2$, $B_1$, $B_2$, $B_3$, $B_4$ and $C$ respectively.
The two-zero texture reduces the number of free parameters
in the neutrino mass matrix, which in term strengthens the predictions
regarding unknown parameters like the Dirac CP phase ($\delta$), the octant degeneracy,
and the absolute mass scale of neutrinos. In this work, first we have
revisited the two-zero texture scenario and have checked whether
all seven experimentally allowed two-patterns have any preference
on the neutrino mass orderings. We have shown our predictions
of these two-zero patterns in $\delta-m$ plane ($m$ being the mass of the
lightest neutrino) and have also demonstrated the detection prospects
through the neutrinoless double beta decay at KamLAND-Zen \cite{KamLAND-Zen:2024eml}.

In the second part of this work, we have shown how naturally
one can get the two-zero patterns in $m_{\nu}$.
For that, we have considered a well motivated scenario like
the Type-II seesaw \cite{Mohapatra:1980yp, Arhrib:2011uy, Chao:2012mx, Dev:2013hka,
Lu:2016ucn, Biswas:2017dxt, Ghosh:2017pxl}
that requires an extra scalar triplet to write a
new Yukawa interaction with the lepton doublets. The neutrino
mass matrix in the minimal Type-II seesaw contains
too many free parameters, and therefore an underlying symmetry among
three lepton flavours is an interesting idea. We have chosen
$L_{\mu}-L_{\tau}$ symmetry \cite{He:1990pn, He:1991qd, Choubey:2004hn, Adhikary:2006rf,
Heeck:2011wj, Park:2015gdo, Biswas:2016yan, Banerjee:2018eaf,
Biswas:2019twf, Jho:2019cxq, Jho:2020sku, Jho:2020jsa}
as the underlying symmetry as it has
other motivations like an anomaly free symmetry without
requiring any additional degree of freedom, successful
explanation of $(g-2)_{\mu}$ anomaly \cite{Ma:2001md, Biswas:2019twf} etc. Here,
$L_{\alpha}$ refers to the lepton number for a particular flavour $\alpha$.
Accordingly, the first generation lepton does not have any
$L_{\mu}-L_{\tau}$ charge while it is $+1($-1$)$ for the
second(third) generation lepton. Under the ${L_{\mu}-L_{\tau}}$
symmetry, the elements of the Majorana mass matrix ($m_{\nu}$)
have five different charges like $0$, $+1$, $-1$, $+2$ and $-2$ respectively.  
Therefore, we need five triplets to generate all the elements
in $m_{\nu}$. Fortunately, we do not required to have all the
elements in the active neutrino mass matrix to be nonzero as
there are lesser number of observables (six) compared to the
number of independent parameters in $m_{\nu}$, which is twelve.  
We have found that the structure of $m_{\nu}$ in the minimal
Type-II seesaw model (only one scalar triplet) with
U(1)$_{L_{\mu}-L_{\tau}}$ symmetry is already ruled-out by the existing neutrino
oscillation data. Our general observation is that a realistic neutrino mass matrix
with an internal flavour structure requires more than two scalar triplets in our model. 
In particular, using three triplets with different ${L_{\mu}-L_{\tau}}$
charges (e.g.\,\,0,\,$\pm 1$,\,$\pm 2$), the two-zero patterns can be
obtained naturally\footnote{In \cite{Grimus:2004az} it has been shown
that the texture C can be generated by three SU(2)$_{\rm L}$ scalar
triplets obeying $\mathbb{Z}_4$ symmetry.}. However, a more careful inspection revels
that only the five textures out of the seven
experimentally allowed patterns can be recreated. The remaining two
patterns, namely $A_1$ and $A_2$, are not possible to achieve
in the ${L_{\mu}-L_{\tau}}$ symmetric Type-II seesaw framework
as more than two independent zeros appear, which is forbidden
by the current neutrino oscillation data. Increasing the number of triplets
to five with ${L_{\mu}-L_{\tau}}$ charges $0,\,+1,\,-1,\,+2,\,-2$
will eventually fill all the elements in the neutrino mass matrix
as mentioned earlier. The number of scalar triplets along with
the corresponding texture of $m_{\nu}$ has been listed
in Table \ref{tab:no_of_triplet} where in the last column
we have mentioned the current phenomenological status
of the corresponding texture. Since the current neutrino
oscillation data can accommodate maximum two independent
zeros in $m_{\nu}$, any texture having more than two
independent zeros is not allowed \cite{Frampton:2002yf}.  
\begin{table}[h!]
\begin{center}
\begin{tabular}{|l|ll|l|}
\hline
\multirow{2}{*}{Number of triplets} & \multicolumn{2}{l|}{~~~~~~~~~~~~~~~Number of independent zeros} 
& \multirow{2}{*}{Phenomenological status} \\ \cline{2-3}
                           & \multicolumn{1}{l|}{~~~At least one triplet is}& {~~~~~~~~~No triplet is}   &           \\
                           & \multicolumn{1}{l|}{uncharged under $L_{\mu}-L_{\tau}$}& {uncharged under $L_{\mu}-L_{\tau}$}
                           &            \\ \hline
                           
~~~~~~~~~~~1 & \multicolumn{1}{l|}{~~~~~~~~~~~~~~~4} & {~~~~~~~~~~~~~~~~~5} &{~~~~~~~~~not allowed}  \\ \hline
~~~~~~~~~~~2 & \multicolumn{1}{l|}{~~~~~~~~~~~~~~~3} & {~~~~~~~~~~~~~~~~~4} &{~~~~~~~~~not allowed}  \\ \hline
~~~~~~~~~~~3 & \multicolumn{1}{l|}{~~~~~~~~~~~~~~~2} & {~~~~~~~~~~~~~~~~~3} &{~~allowed/ not allowed} \\ \hline
~~~~~~~~~~~4 & \multicolumn{1}{l|}{~~~~~~~~~~~~~~~1} & {~~~~~~~~~~~~~~~~~2}& {~~~~~~~~~~~~allowed}    \\ \hline
~~~~~~~~~~~5 & \multicolumn{1}{l|}{~~~~~~~~~~~~~~~0} & {~~~~~~~~~~~~~~~N.A.}&{~~~~~~~~~~~~allowed}    \\ \hline
\end{tabular}
\end{center}
\caption{Number of triplets and corresponding number of
independent zeros in the neutrino mass matrix.}
\label{tab:no_of_triplet}
\end{table}


{The multiple scalar triplets can be found naturally in many
BSM scenarios with larger symmetry groups like the Left-Right symmetric model
(${\rm SU(3)}_{\rm c} \times {\rm SU(2)}_{\rm L} \times {\rm SU(2)}_{\rm R} \times
{\rm U(1)}_{\rm B-L}$) \cite{Mohapatra:1974hk, Mohapatra:1974gc, Grimus:1993fx, Garcia-Cely:2015quu},
the $331$ model (${\rm SU(3)}_{\rm c} \times {\rm SU(3)}_{\rm L} \times
{\rm U(1)}_{\rm X}$) \cite{Long:1997vbr, Byakti:2020ipa} etc.
Although the presence of multiple triplets in the scalar
potential may complicate the scenario, it has some advantages
over the single triplet in the context of detection at collider.
One of the interesting features of scalar triplet
is the presence of a doubly charged scalar $\Delta^{++}$ in the spectrum.
In the Type-II seesaw model with only one scalar triplet, the
doubly charged scalar has only two prominent decay modes depending
on the value of the triplet vacuum expectation value (VEV). For example, $\Delta^{++}$
predominantly decays into same sign lepton pairs ($\Delta^{++}\rightarrow \ell^{+}\ell^{+}$)
when VEV $\lesssim {\rm 10^{-4}}$ GeV while $\Delta^{++}\rightarrow W^{+}W^{+}$ becomes dominant
for VEV $> {\rm 10^{-4}}$ GeV.  However, the presence of multiple
triplets introduces another unique decay mode where a doubly
charged scalar in one triplet $\Delta^{++}_i$ decays into a
singly charged scalar $\Delta^{+}_j$ of another triplet accompanied
by the SM $W^{+}$ boson ($\Delta_i^{++}\rightarrow \Delta_j^{+}W^{+}$, $i\neq j$)
\cite{Chaudhuri:2013xoa}, which is otherwise kinematically forbidden
for the case of a single triplet due to small mass splitting among the members of a
scalar triplet resulting from the electroweak precision test \cite{Chun:2012jw, Ghosh:2017pxl}. 
Additionally, $\Delta^{++}_{i} \rightarrow \Delta^{++}_{j} h$ is also a
viable decay mode which eventually produces final states like
$\ell^{+}\ell^{+}+2\,\nu_{\ell} + h$ and
$\ell^{+}\ell^{+}+h$ respectively depending on the triplet
VEV \cite{Chaudhuri:2016rwo, KumarGhosh:2018bli}. Moreover, a rich scalar sector,
such as the present scenario, can give rise to a stochastic gravitational
wave background originating from a strong first-order phase transition,
which we have studied in detail in \cite{Biswas2025xp}.}

Rest of the paper is organised as follows.
In Section\,\ref{sec:twozero}, we introduce two-zero texture and
different two-zero patterns allowed by the experiments.
The numerical results on the neutrino
oscillation parameters for two-zero texture are shown
in Section\,\ref{sec:results}. Implementation of
two-zero texture in Type-II seesaw model has been
discussed in Section\,\ref{sec:typeII}. The Section\,\ref{sec:constaints}
is dedicated to the existing constrains on the Yukawa couplings
from lepton flavour violating rare decays. Finally, we
present our conclusion in Section\,\ref{sec:conclusion}.
\section{Two-zero texture}
\label{sec:twozero}
From the neutrino oscillation experiments, we know three mixing angles
($\theta_{12}$, $\theta_{23}$ and $\theta_{13}$) and two mass-square
differences\footnote{mass-square difference between $i$th
and $j$th mass eigenstate is defined as $\Delta{m^{2}_{ij} = m^2_i - m^2_j}$}
($\Delta{m^2_{21}}$ and $\Delta{m^2_{31}}$) precisely except the sign of
the quantity $\Delta{m^2_{31}}$ which can be positive or negative
depending on the mass hierarchy of the neutrinos
(normal hierarchy (NH) or inverted hierarchy (IH))
\footnote{for the normal hierarchy $m_3>> m_2 \gtrsim m_1$ while
for the inverted case $m_2 \gtrsim m_1 >> m_3$}. Besides, we have
a poor understanding about CP violation in the leptonic sector. The
recent results \cite{Esteban:2024eli} using experimental data indicate a large
uncertainty for the Dirac CP phase ($\delta$) in $3 \sigma$
range between $108\degree$ and $404\degree$ ($192\degree$ to $360\degree$)
for NH (IH). However, the actual
flavour structure of the neutrino mass matrix is still unknown. There
are numerous possibilities that can reproduce six experimental
observables ($\theta_{12}$, $\theta_{23}$, $\theta_{13}$, $\Delta{m^2_{21}}$,
$\Delta{m^2_{31}}$ and $\delta$) since the neutrino mass matrix, being a
$3\times 3$ complex symmetric matrix (if neutrinos are Majorana fermions),
has six independent complex elements (or twelve real elements). Therefore,
more than one elements can be vanishingly small or even exactly
zero also due to an underlying symmetry between different neutrino flavour
eignstates. It has been shown in \cite{Fritzsch:2011qv} that the present neutrino
oscillation data allow only two independent zeros in the neutrino
mass matrix and there are fifteen possible two-zero textures
since $^{6}\mathbb{C}_{2} = \dfrac{6!}{4!~2!} = 15$. Out of
these fifteen textures, only seven are capable of generating
the correct mass square differences and mixing angles and
these are given below:
\begin{eqnarray}
&&{A_1} = \begin{pmatrix}
0 & 0 & \times \cr 0 & \times &
\times \cr \times & \times & \times \cr 
\end{pmatrix}, \,\,
A_2 = 
\begin{pmatrix}
 0 & \times & 0 \cr \times & \times &
\times \cr 0 & \times & \times \cr
\end{pmatrix},\,\, \nonumber \\
&&B_1 = 
\begin{pmatrix}
 \times & \times & 0 \cr \times & 0 &
\times \cr 0 & \times & \times \cr
\end{pmatrix}, \,\, 
B_2 = 
\begin{pmatrix}
 \times & 0 & \times \cr 0 & \times & \times \cr
\times & \times & 0 \cr
\end{pmatrix},\,\,
B_3 = \begin{pmatrix}
\times & 0 & \times \cr 0 & 0 &
\times \cr \times & \times & \times \cr
\end{pmatrix},\,\,
B_4 = \begin{pmatrix}
 \times & \times & 0 \cr \times & \times &
\times \cr 0 & \times & 0 \cr
\end{pmatrix} ,\nonumber \\
&&C = \begin{pmatrix}
\times & \times & \times \cr \times & 0 &
\times \cr \times & \times & 0 \cr
\end{pmatrix}\,\,,
\label{eq:texture}
\end{eqnarray}
where the $\times$ symbol indicates
nonzero element in the mass matrix. Moreover,
if analyses these seven textures more critically,
one can find that all the seven types are not entirely independent. 
Some of the textures are related by a permutation symmetry and hence
possess identical characteristics. For example, the texture ${\rm A}_2$
can be obtained from the texture ${A}_1$ by interchanging
between the second and the third columns and the second
and the third rows respectively. Mathematically, this
transformation can be expressed as
\begin{eqnarray}
{A}_2 = P^{T}_{23}\,\,{A}_1\,P_{23}\,, 
\end{eqnarray}   
where $P_{23} = \left(\begin{smallmatrix}
1~&~0~&~0 \\ \\ 0~&~0~&~1 \\ \\ 0~&~1~&~0
\end{smallmatrix}\right)$ is an orthogonal matrix. Similarly,
the textures $B_1$, $B_2$ and $B_3$, $B_4$, respectively, are also
related by the permutation symmetry as described above. Therefore,
in principle, we have only four distinct types of two-zero textures
namely $A_1$, $B_1$, $B_3$ and $C$ which are currently allowed
by the neutrino oscillation data. We would like to note
that although there are four complex elements
(or equivalent to eight real parameters) in $m_{\nu}$ (Eq.\,\eqref{eq:texture}), one can
further reduce three parameters using phase rotation
freedom of the lepton fields. As a result,
a two-zero texture can be described effectively by five
parameters only (three real and two complex parameters).

If $m_\nu$ is the neutrino mass matrix in the flavour basis
and $m^{dia}_{\nu} = {\rm dia}\left(m_1, m_2, m_3\right)$
is the diagonalised neutrino mass matrix in the mass basis,
then
\begin{eqnarray} 
m^{\rm dia}_{\nu} = V^{T} m_{\nu} V\,,
\label{eq:mdia}
\end{eqnarray}
 where $V= U_{\rm PMNS} P$ is a 
product of the PMNS matrix and a phase matrix $P$, which are given by 
\begin{eqnarray}
U_{\rm PMNS} = 
\begin{pmatrix}
c_{13}c_{12} & c_{13}s_{12} & s_{13}e^{-i\delta}\cr
-c_{23}s_{12}-s_{23}s_{13}c_{12}e^{i\delta} & c_{23}c_{12}-s_{23}s_{13}s_{12}e^{i\delta} & s_{23}c_{13}\cr
s_{23}s_{12}-c_{23}s_{13}c_{12}e^{i\delta} & -s_{23}c_{12}-c_{23}s_{13}s_{12}e^{i\delta} &
c_{23}c_{13}\cr
\end{pmatrix}
~{\rm and}~
P = {\rm dia}(e^{i\rho}, e^{i\sigma}, 1)\,\,,
\nonumber \\
\label{eq:U&P}
\end{eqnarray}
where two additional phases $\rho$ and $\sigma$ are
known as Majorana phases. Now, using the definition
of phase matrix $P$, the neutrino mass matrix
in the flavour basis can be written as
\begin{eqnarray}
m_{\nu} = U^{*}_{\rm PMNS}~\Lambda~U^{\dagger}_{\rm PMNS}\,, 
\end{eqnarray}
where $\Lambda = P^{*}\,m^{\rm dia}_{\nu}\,P^\dagger = {\rm dia}(m_1 e^{-2i\rho},\,\,m_2 e^{-2i\sigma},\,\,m_3)$.
Therefore, for a two-zero texture having two
independent zeros ($\alpha\,\beta$ and $a\,b$ elements), one can have
\begin{eqnarray}
&& \sum_i (U^*_{\rm PMNS})_{\alpha\,i} \Lambda_{i\,i} (U^\dagger_{\rm PMNS})_{i\,\beta} = 0 
\label{eq:albt} \\
&& \hspace{-4cm}{\rm and} \nonumber \\
&& \sum_j (U^*_{\rm PMNS})_{a\,j} \Lambda_{j\,j} (U^\dagger_{\rm PMNS})_{j\,b} = 0 \,\,.
\label{eq:ab}
\end{eqnarray}   
Solving Eqs.\,(\ref{eq:albt} and \ref{eq:ab}), we can express
the ratios of $\Lambda_{11}/\Lambda_{33}$ and $\Lambda_{22}/\Lambda_{33}$
in terms of the elements of the PMNS matrix as  
\begin{eqnarray}
\dfrac{\Lambda_{11}}{\Lambda_{33}} &=& 
\dfrac{W_{a 2} W_{b 2} W_{\alpha 3} W_{\beta 3} - W_{a 3}W_{b 3}W_{\alpha 2}W_{\beta 2}}
{W_{\alpha 2}W_{\beta 2} W_{a 1}W_{b 1} - W_{\alpha 1}W_{\beta 1}W_{a 2}W_{b 2}}\,,
\label{eq:lam1}\\
\dfrac{\Lambda_{22}}{\Lambda_{33}} &=& 
\dfrac{W_{a3} W_{b3} W_{\alpha 1} W_{\beta 1} - W_{a1}W_{b1}W_{\alpha 3}W_{\beta 3}}
{W_{\alpha 2}W_{\beta 2} W_{a 1}W_{b 1} - W_{\alpha 1}W_{\beta 1} W_{a2}W_{b2}}\,,
\label{eq:lam2}
\end{eqnarray}
where the matrix $W = U^*_{\rm PMNS}$ and $\Lambda_{11} = m_1 e^{-2i\rho}$,
$\Lambda_{22} = m_2 e^{-2i\sigma}$, $\Lambda_{33} = m_3$ respectively. Therefore,
the mass eigenvalues $m_1$ and $m_2$ can easily be obtained from
Eqs.\,\,(\ref{eq:albt} and \ref{eq:ab}) for $m_3$ and other experimental
observables. The Majorana phases $\rho$ and $\sigma$ are given by
\begin{eqnarray}
\rho = -\frac{1}{2} {\rm arg}\left(\frac{\Lambda_{11}}{\Lambda_{33}}\right),~~~
\sigma = -\frac{1}{2} {\rm arg}\left(\frac{\Lambda_{22}}{\Lambda_{33}}\right)\,.
\label{eq:majo_phases}
\end{eqnarray}
\section{Numerical results: masses and mixing angles}
\label{sec:results}
\begin{figure}
\subfigure[Texture $A_1$]{
\includegraphics[height=5cm,width=8cm]{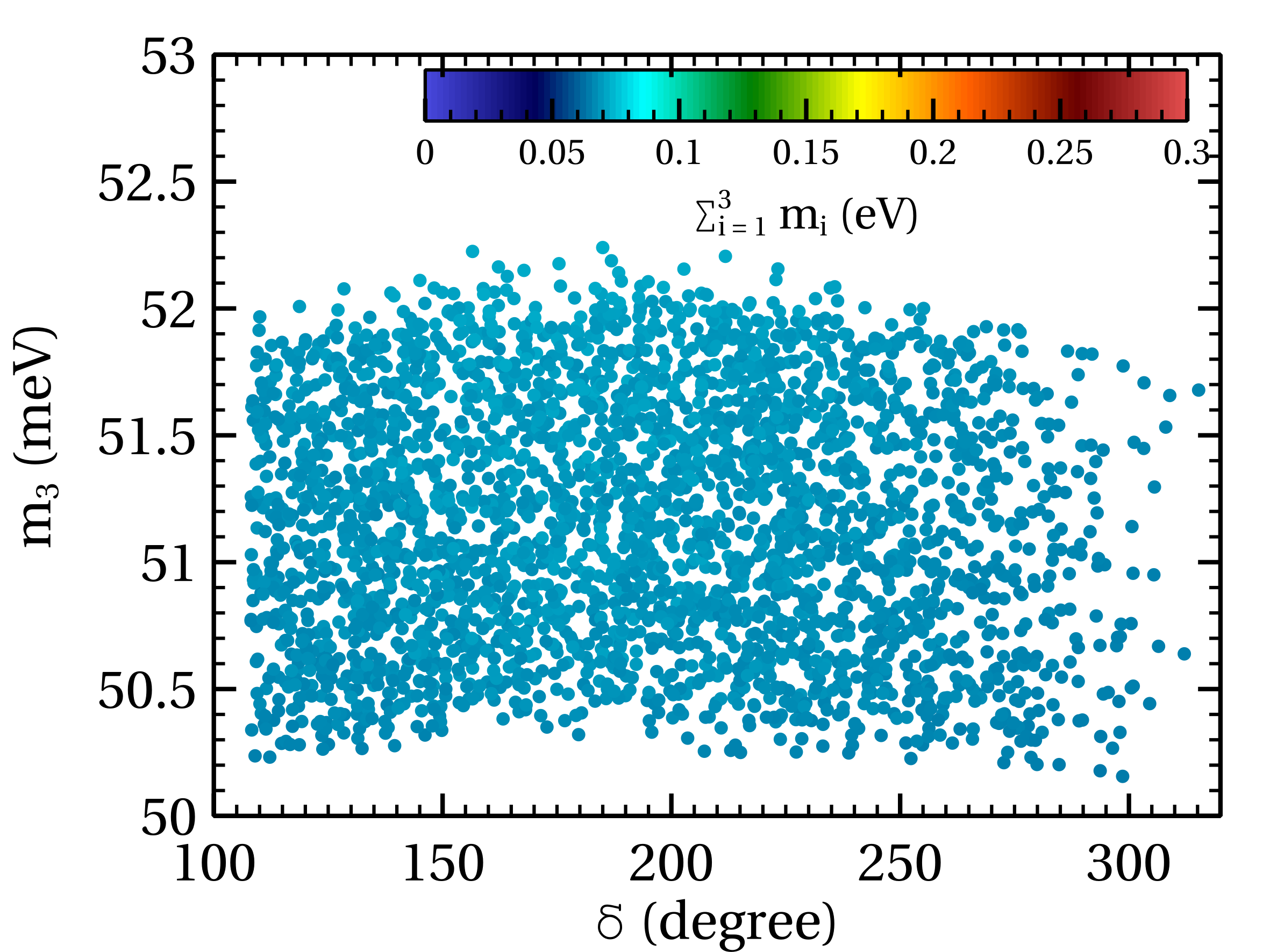}}
\subfigure[Texture $A_2$]{
\includegraphics[height=5cm,width=7.6cm]{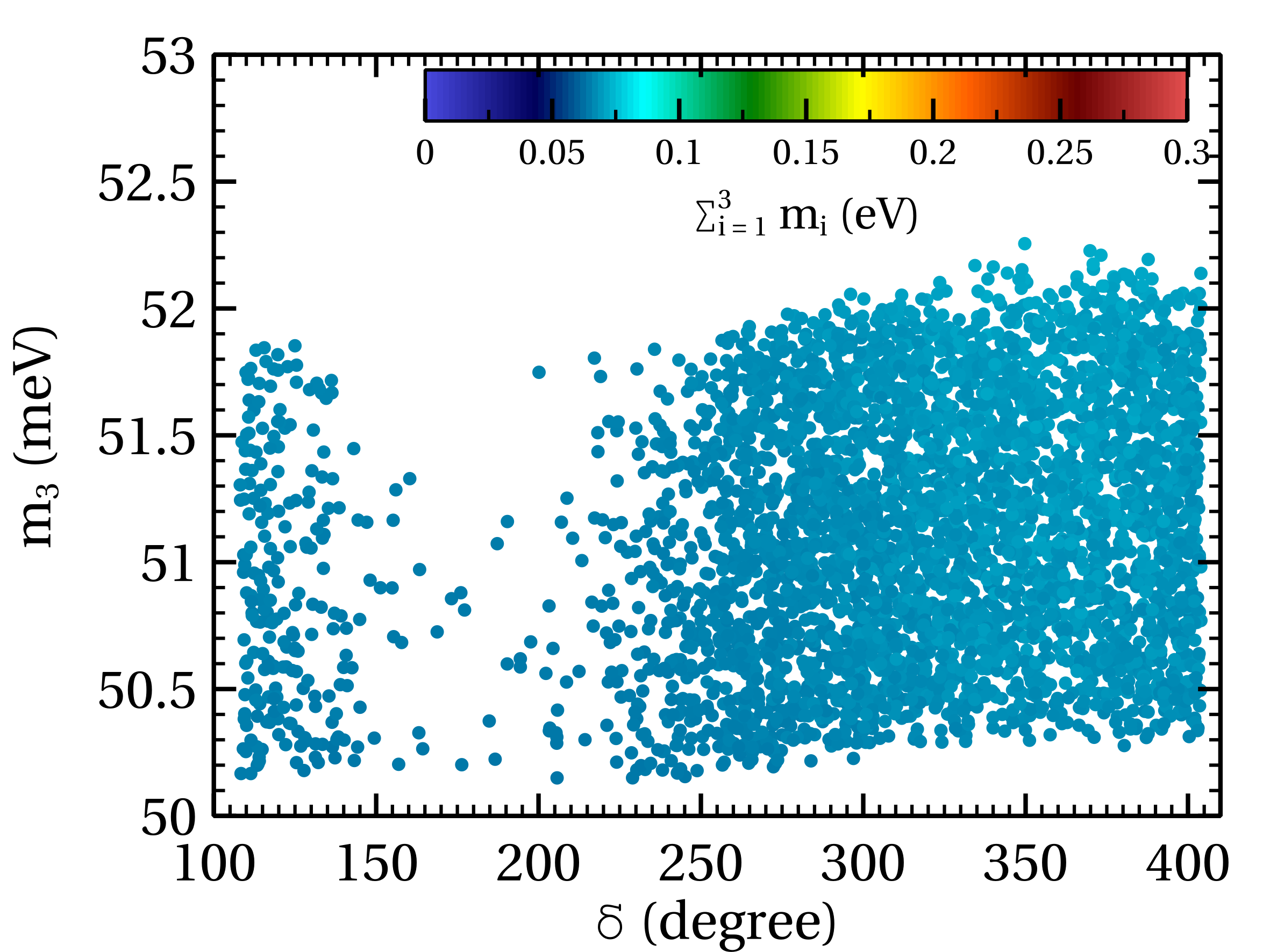}}\\
\subfigure[Texture $B_1$]{
\includegraphics[height=5cm,width=7.5cm]{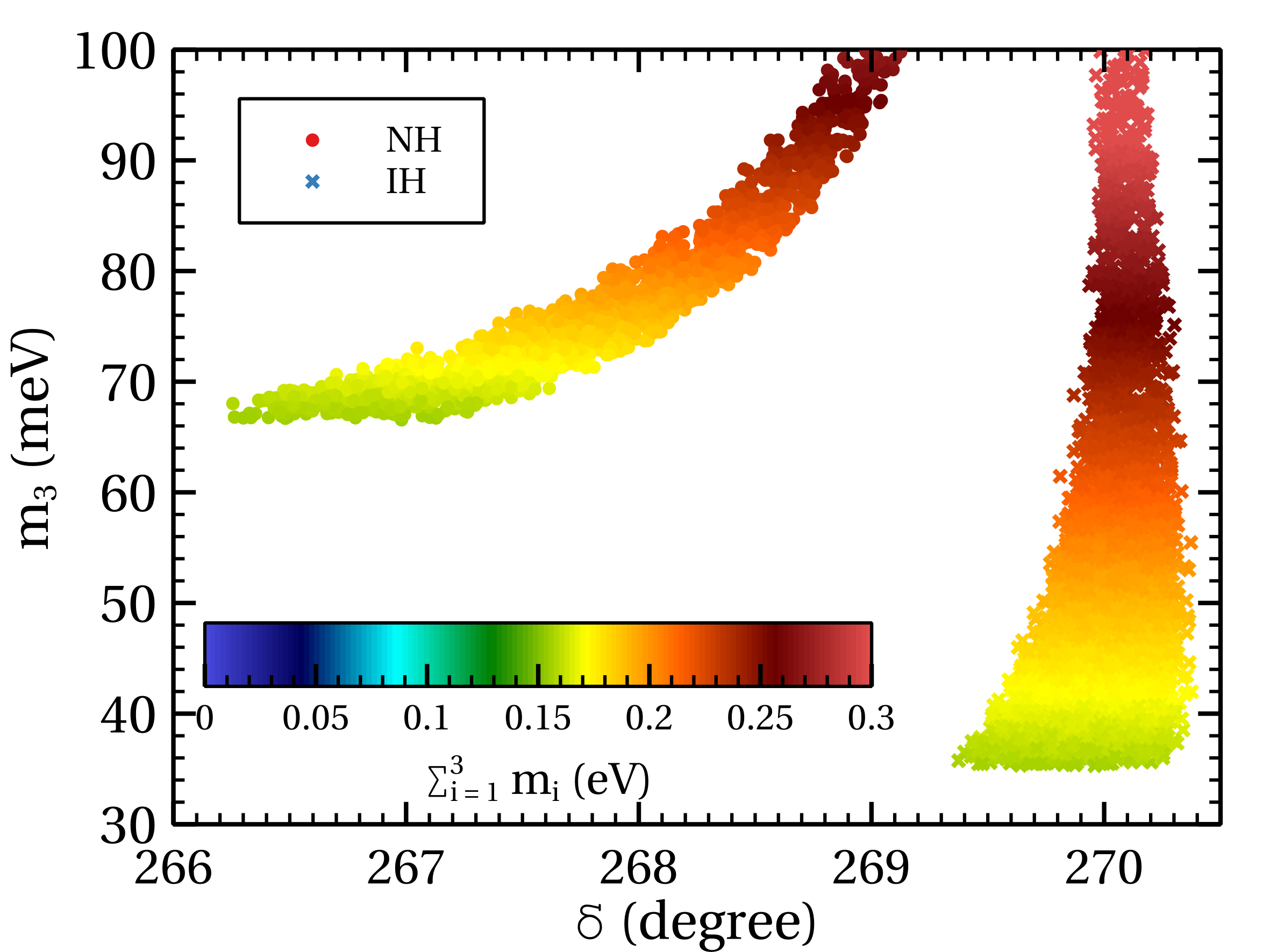}}
\subfigure[Texture $B_2$]{
\includegraphics[height=5cm,width=7.5cm]{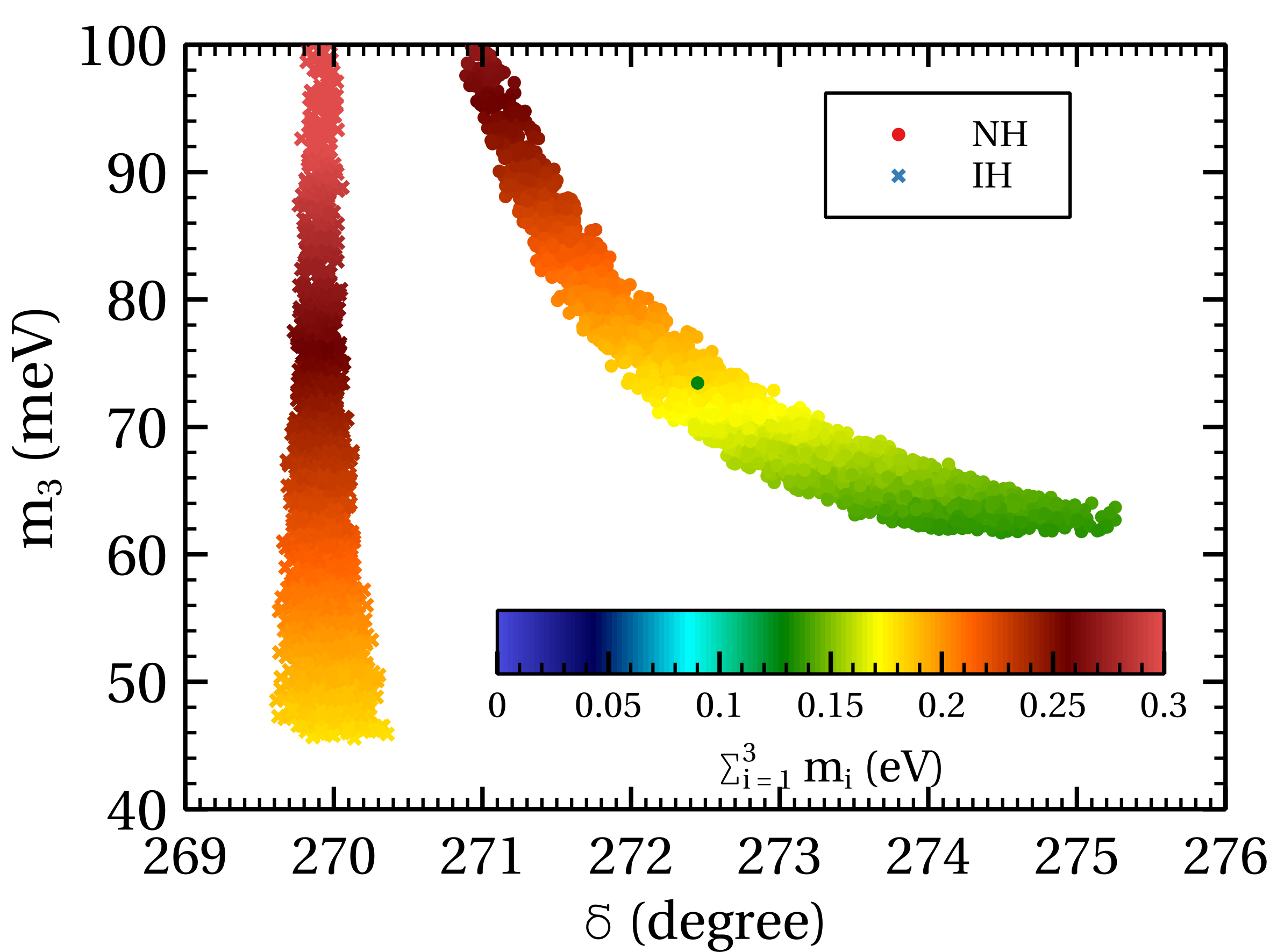}}\\
\subfigure[Texture $B_3$]{
\includegraphics[height=5cm,width=7.5cm]{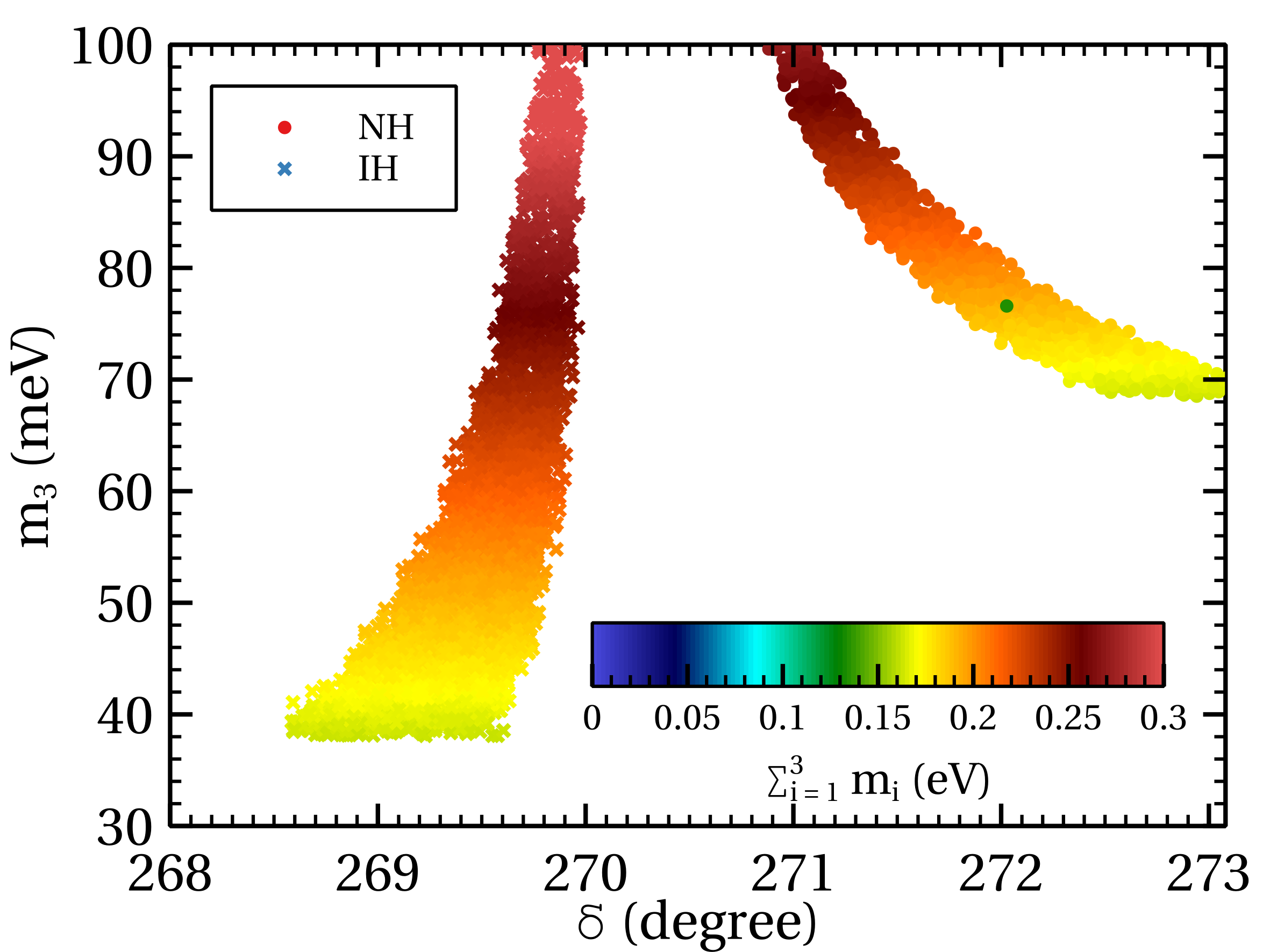}}
\subfigure[Texture $B_4$]{
\includegraphics[height=5cm,width=7.5cm]{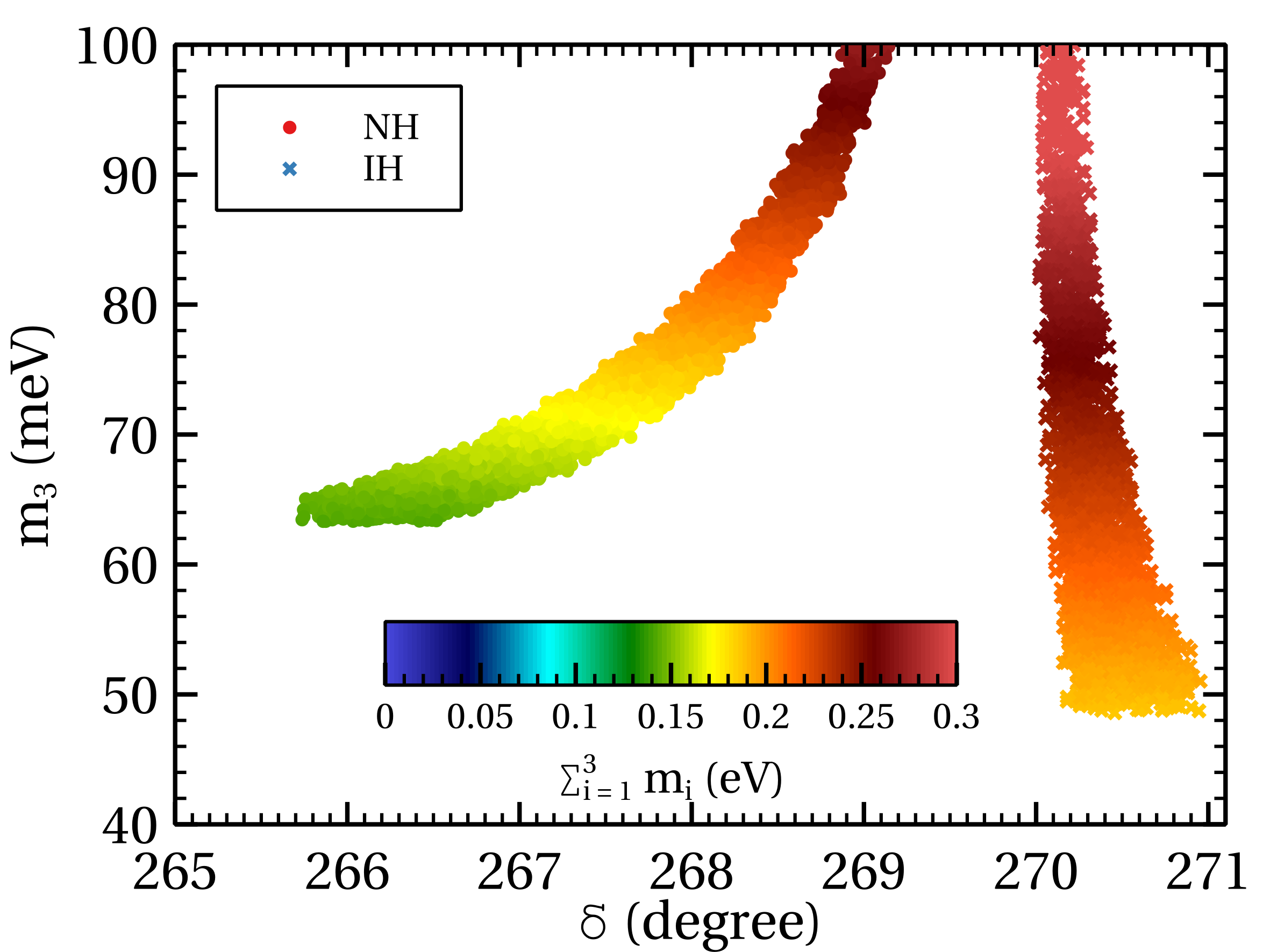}}\\
\subfigure[Texture $C$]{
\includegraphics[height=5cm,width=7.5cm]{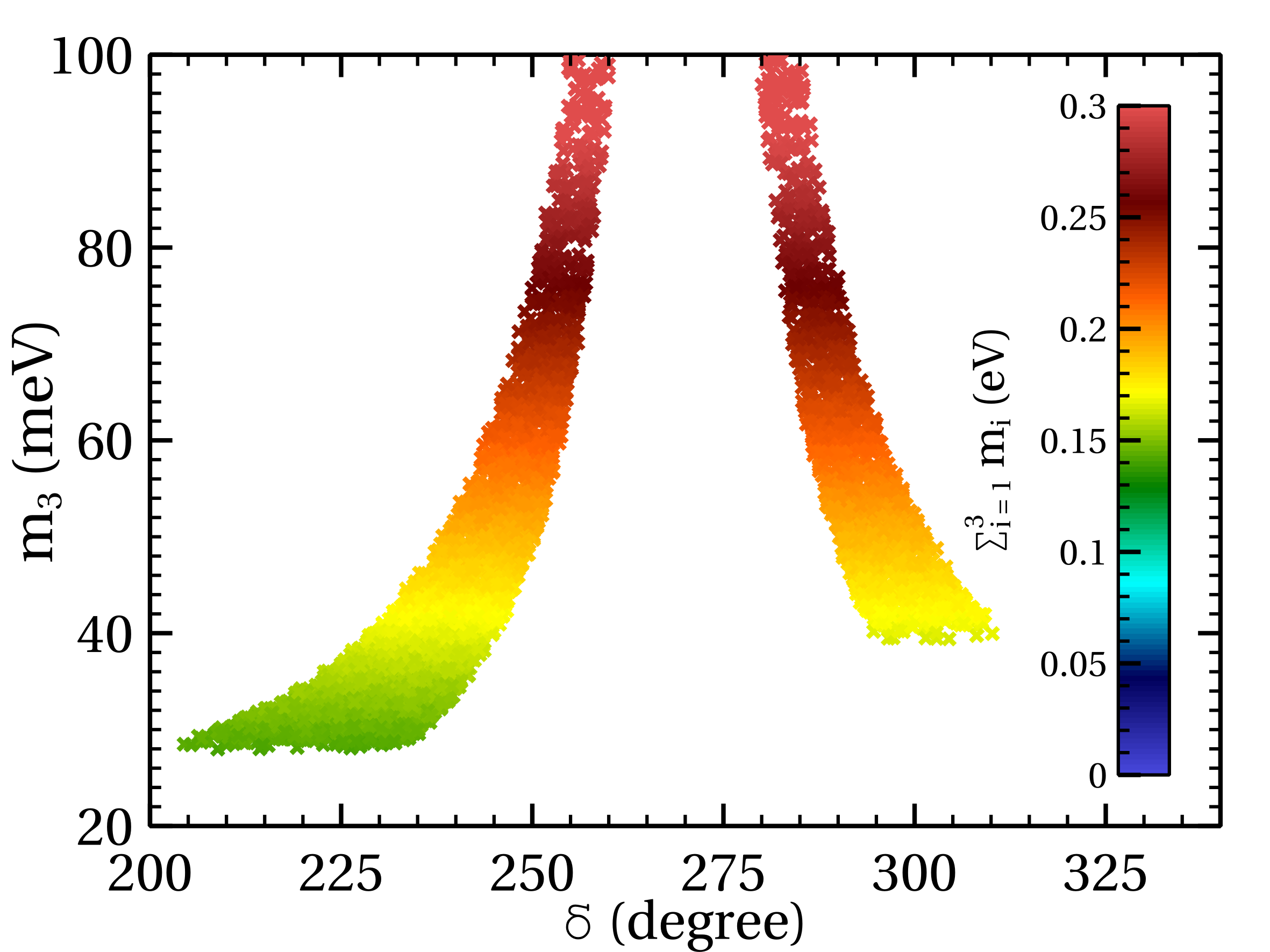}}
\caption{$\delta-m_3$ plane for all allowed two-zero textures.}
\label{fig:delta-m3}
\end{figure}

In this section, we have presented our results
on neutrino oscillation parameters for the two-zero
texture described earlier. We have computed mass eigenvalues
for each allowed two-zero texture (Eq.\,\eqref{eq:texture}) using
Eqs.\,\eqref{eq:lam1} and \eqref{eq:lam2} numerically. For this
we have taken the latest $3\sigma$ range allowed values \cite{Esteban:2024eli}
of three mixing angles ($\theta_{12}$, $\theta_{23}$ and $\theta_{13}$)
and the Dirac CP phase $\delta$, which are appearing in the
right hand side of Eqs.\,\eqref{eq:lam1} and \eqref{eq:lam2}
as the elements of PMNS matrix. Finally, a trial value of
$m_3$ is required to compute the other two mass eigenvalues ($m_1$
and $m_2$) from the mass two mass ratios. The trial value of
$m_3$ is allowed by the oscillation data only if the two mass square
differences $\Delta{m^2_{21}}$ and $\Delta{m^2_{3\ell}}$ remain within
the $3\sigma$ range as determined from neutrino oscillation
experiments. The latest values of mass square differences
and mixing angles for both normal and inverted mass
hierarchies are listed in Table \ref{tab:oscillation_data}.

In Fig.\,\ref{fig:delta-m3}, we have shown the $\delta-m_3$
plane using Eqs.\,(\ref{eq:lam1} and \ref{eq:lam2}) for
the allowed seven textures given in Eq.\,\eqref{eq:texture}.
In each plot, all the points satisfy neutrino oscillation data, 
i.e. two mass square differences and three mixing angles in the $3\sigma$ range.
The colour bar indicates the sum of all three neutrino masses in eV scale.
We have found that for the textures $A_1$ and $A_2$, (Fig.\,\ref{fig:delta-m3}(a)
and \ref{fig:delta-m3}(b)), the normal mass ordering is the only possibility
with the largest mass eigenstate $m_3$ lies in the range $0.05\,\,{\rm eV} <
m_{3}\lesssim 0.052$ eV. The corresponding Dirac CP phase $\delta$ is uniformly
distributed between $110\degree - 300\degree$ and a few points with
$\delta > 300\degree$ for the texture $A_1$. However,
for $A_2$, we have two distinct regions in $\delta-m_3$ plane
for $110\degree\lesssim \delta \lesssim 150$ and $200\degree \lesssim \delta \lesssim
400\degree$ respectively. On the other hand, both normal and inverted
mass orderings can be realised for B textures and we have obtained specific
range for the Dirac CP phase $\delta$. For example,
in the case of normal hierarchy, the Dirac CP phase
lies between $266\degree - 269\degree$ with the corresponding
$m_3$ lies in the range $0.065\,\,{\rm eV} \lesssim m_3
\lesssim 0.1\,\,{\rm eV}$ for textures $B_2$ and $B_3$. For the other two textures $B_1$
and $B_4$, the prediction for $\delta$ is above $270\degree$
(for NH). In the case of inverted mass ordering, the CP
phase in all four cases lies around $270\degree$ along with
$0.04\,\,{\rm eV}\lesssim m_3 \lesssim 0.1$ eV. Finally,
for texture $C$, only the inverted hierarchy is
allowed\footnote{For the normal hierarchy, oscillation
parameters are satisfied only when $m_3>0.16$ eV resulting in
a quasi degenerate mass spectrum with the sum of three neutrino
masses is at least four times larger than the current
cosmological bound \cite{Planck:2018vyg}.}
and here we have found two distinct bands
for $\delta$ in either side of $270\degree$.

Moreover, in all the plots in Fig.\,\,\ref{fig:delta-m3}, the 
sum of three light neutrino masses ($ \sum_{n=1}^{3} m_i $)
is indicated by the colour bar.
One of the important observations that we have noticed in
Fig.\,\,\ref{fig:delta-m3} that expect, for the textures
$A_1$ and $A_2$, the sum of three light neutrino masses
lies in the range $\sim$  0.15 meV $-$ 0.2 meV for all
other textures. This value is slightly larger than the present 
limit $\sum_{n=1}^{3} m_i\leq 0.12$ eV obtained from CMB experiment
by the Planck collaboration \cite{Planck:2018vyg}. Actually, the upper
bound on the sum of neutrino masses comes from the constraint
on neutrino contribution to the total energy budget of the Universe,
which is $\Omega_{\nu} h^2 \equiv \dfrac{\sum m_i n^0_{\nu}}
{\rho_{c}} h^2 < 1.3\times10^{-3}$ \cite{Escudero:2022gez}, where $n^0_{\nu}$
is the present number density of neutrinos per helicity
state and according to the SM, $n_{\nu}^0= 56.9/{\rm cm}^3$
which is calculated assuming neutrinos continue
to follow the Fermi-Dirac distribution with a
different temperature ($T_{\nu} = (4/11)^{1/3} T_{\gamma}$)
than the photon temperature $T_{\gamma}$ after
decoupling at $T_{\gamma} \simeq 1$ MeV. Since
the only constraint is on the relic density of
neutrinos (i.e. the product of mass and number density),
it is possible to relax the bound on sum of neutrino
masses if the number density of neutrinos reduce
from the value predicted by the SM. One of the
ways to reduce the neutrino number density is considering
a scenario where neutrino decays into new lighter BSM states after
decoupling. The detailed procedure can be found in \cite{Farzan:2015pca,Escudero:2022gez}
and by introducing several new states the authors have shown that one
can accommodate the sum of neutrino masses as large as 1 eV.   

\begin{figure}
\subfigure[Texture $A_1$]{
\includegraphics[height=5cm,width=7.5cm]{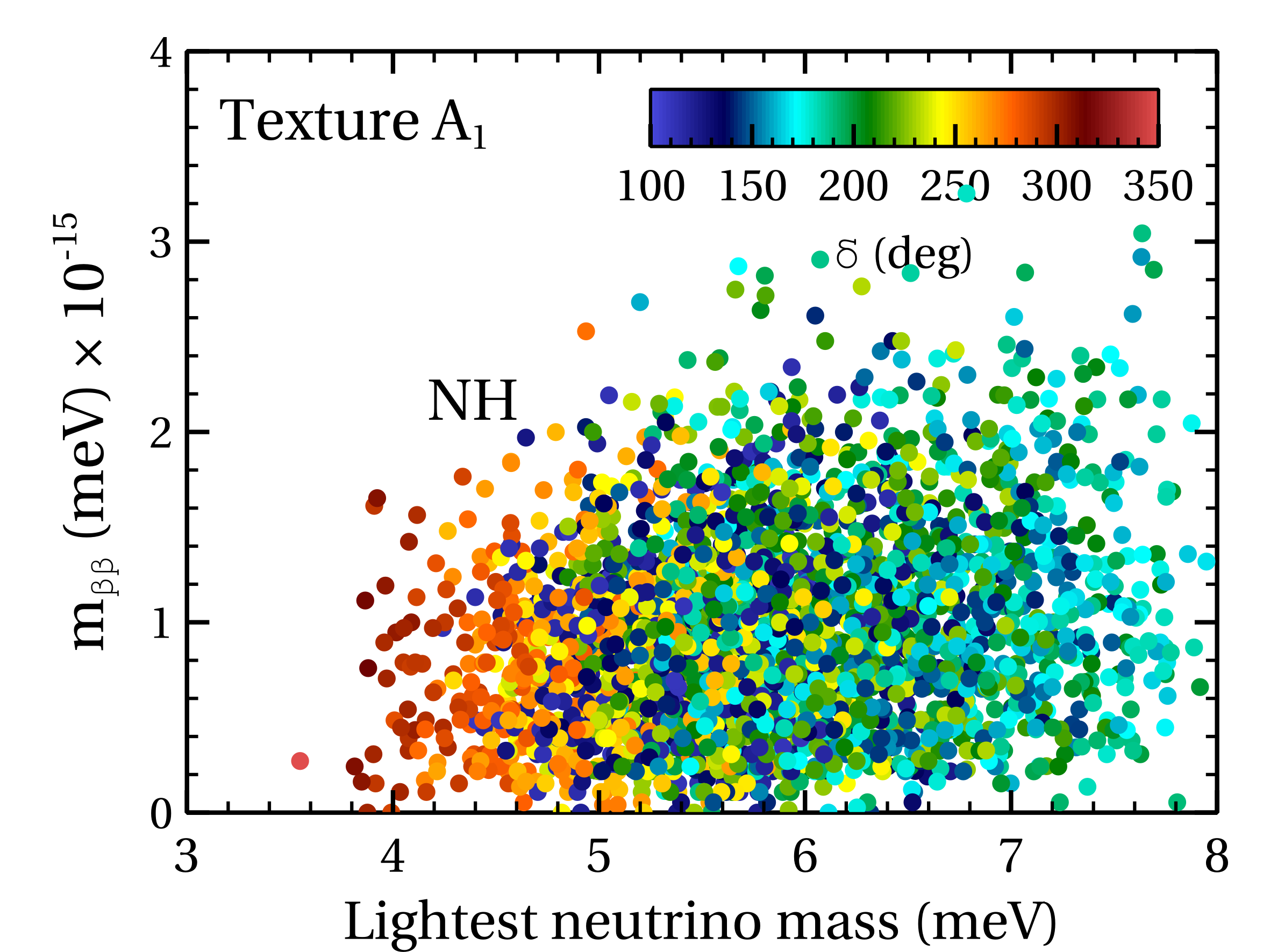}}
\subfigure[Texture $A_2$]{
\includegraphics[height=5cm,width=7.5cm]{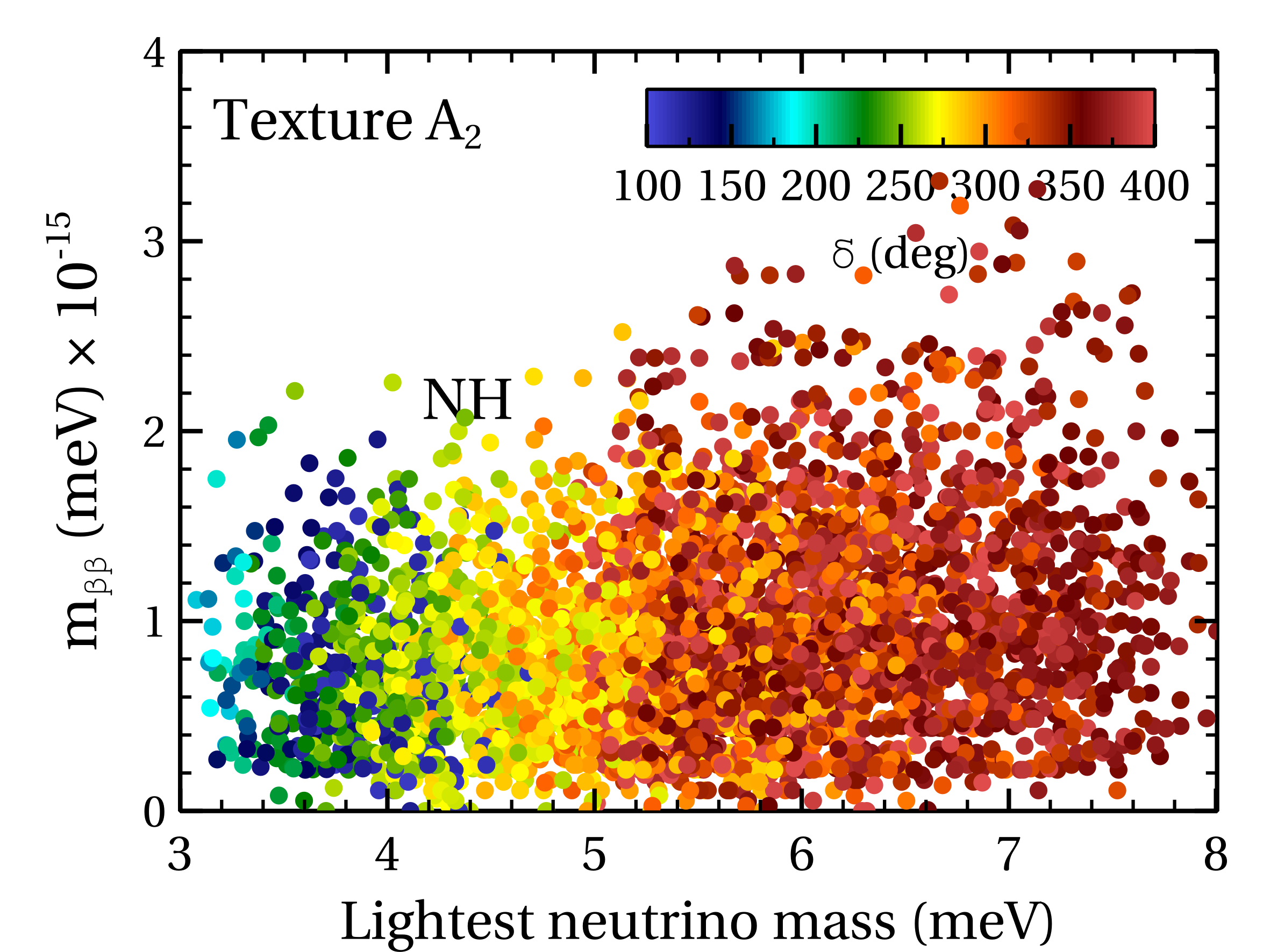}}\\
\subfigure[Texture $B_1$]{
\includegraphics[height=5cm,width=7.5cm]{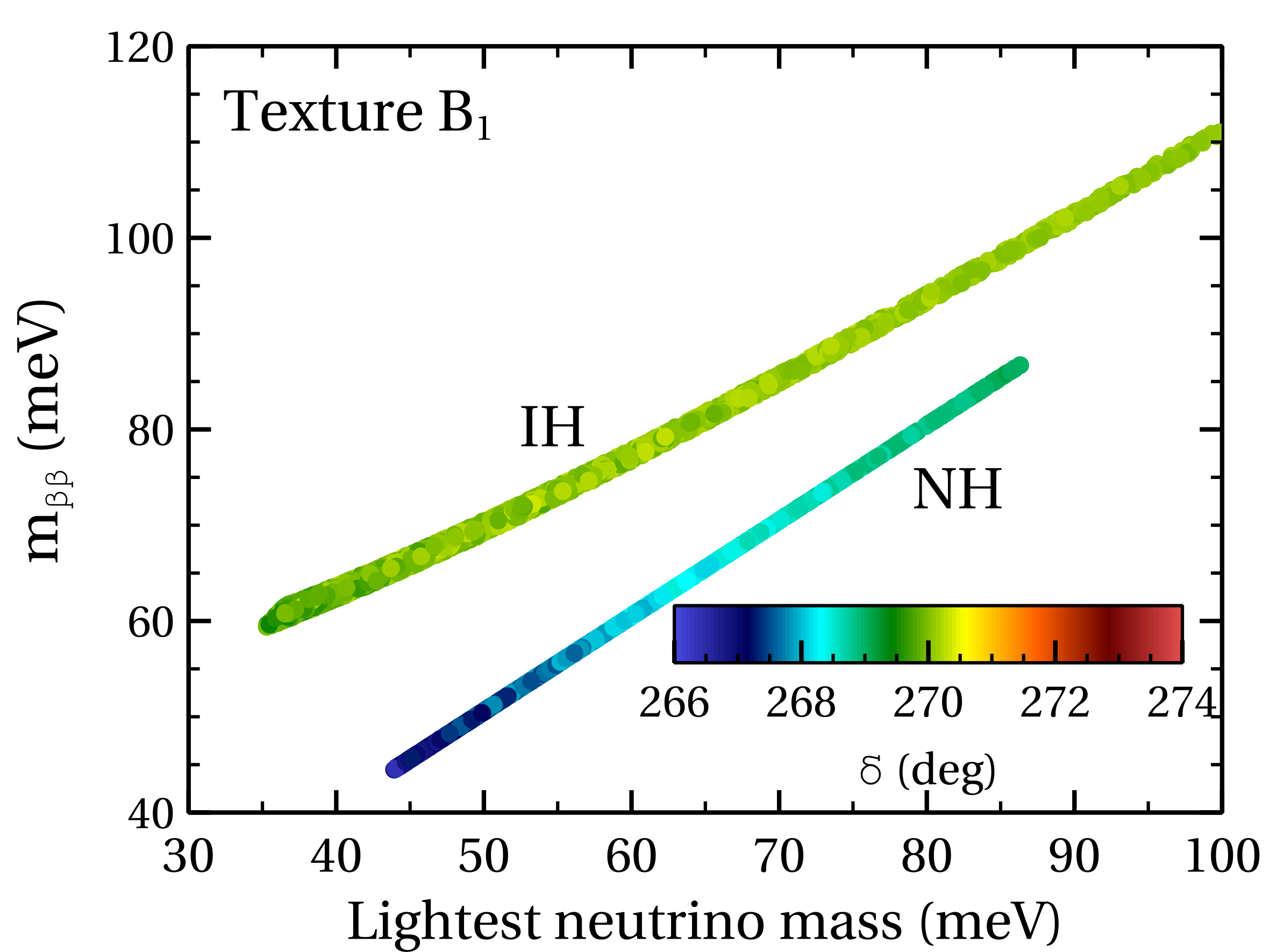}}
\subfigure[Texture $B_2$]{
\includegraphics[height=5cm,width=7.5cm]{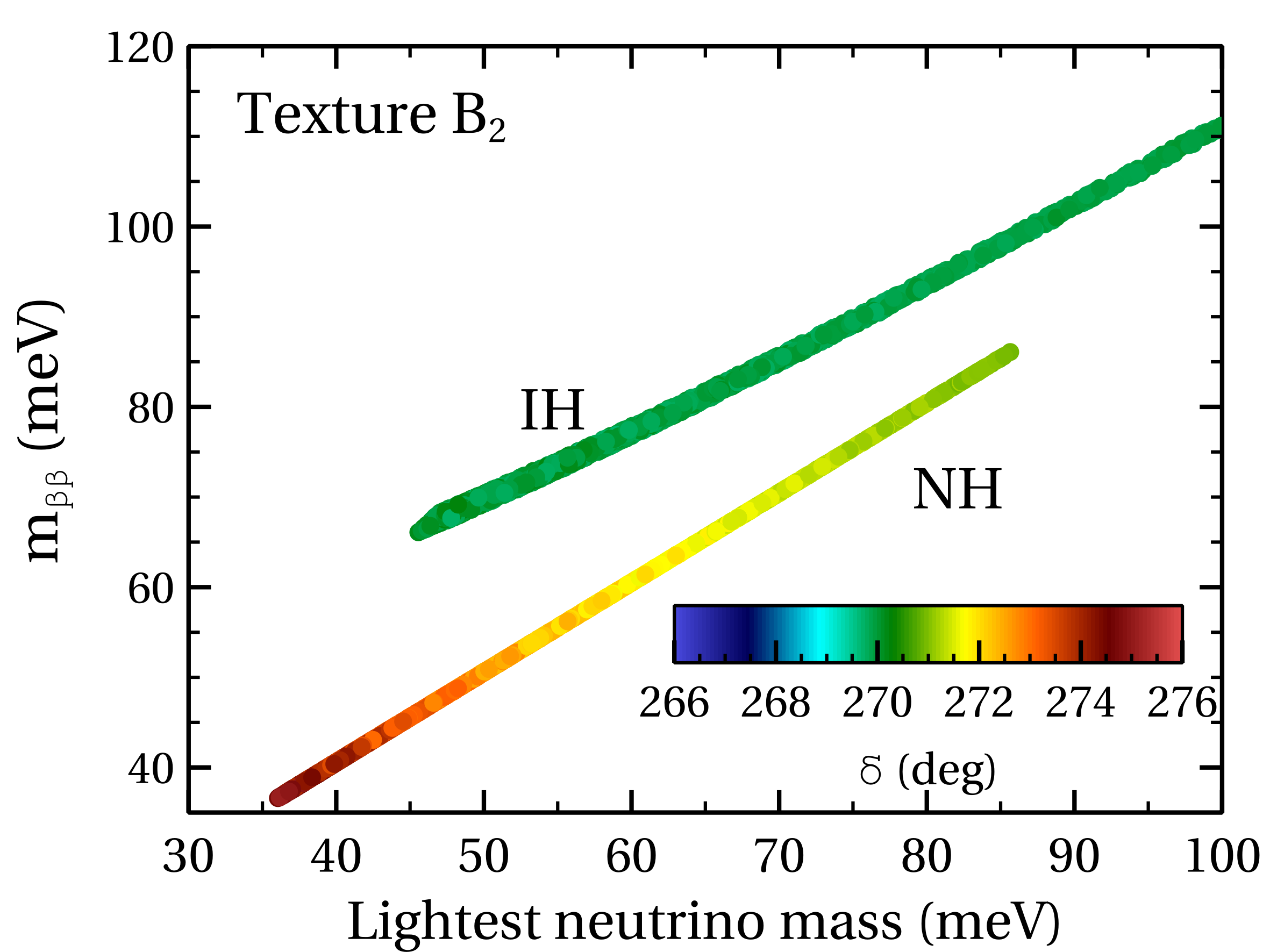}}\\
\subfigure[Texture $B_3$]{
\includegraphics[height=5cm,width=7.5cm]{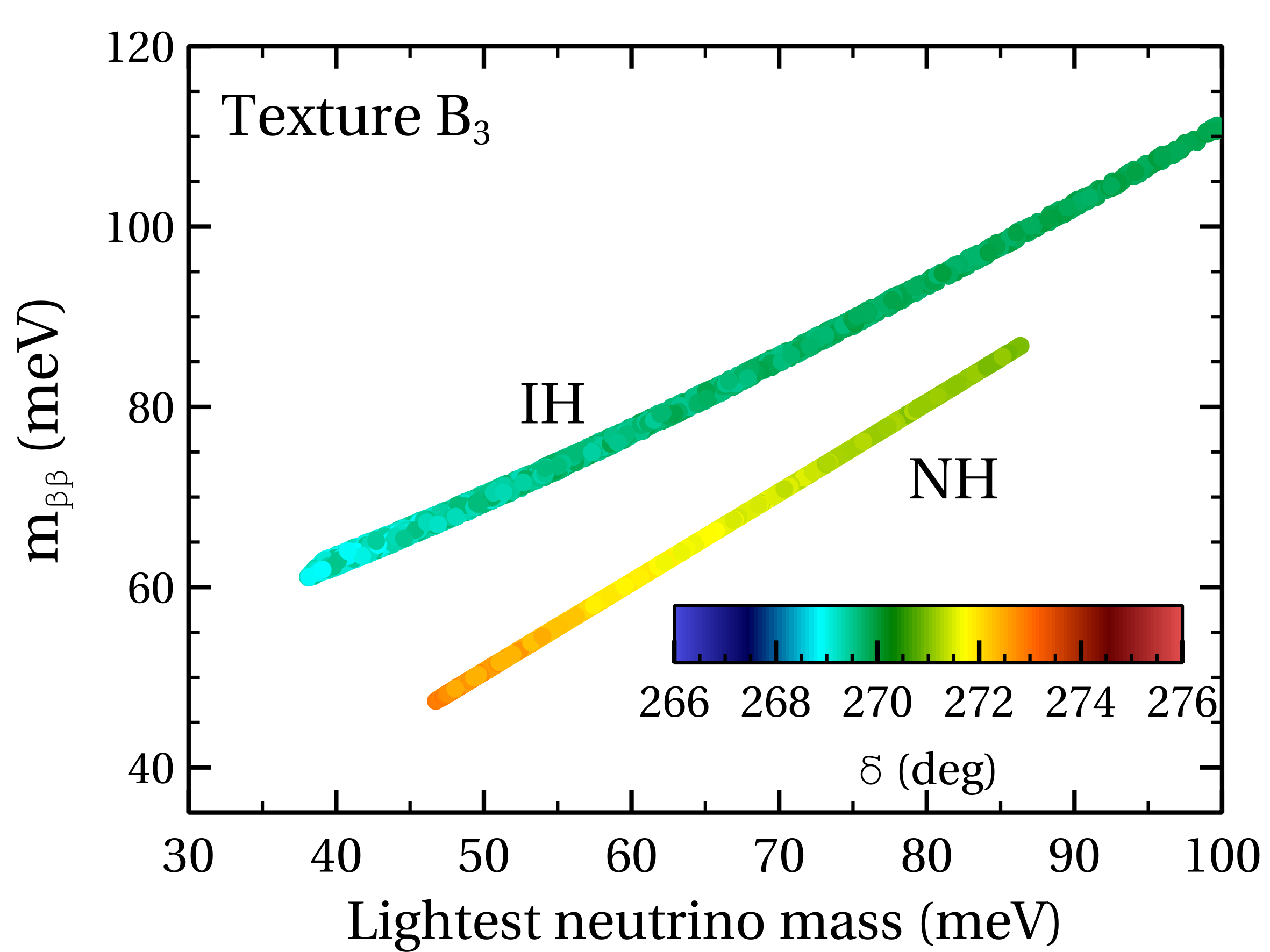}}
\subfigure[Texture $B_4$]{
\includegraphics[height=5cm,width=7.5cm]{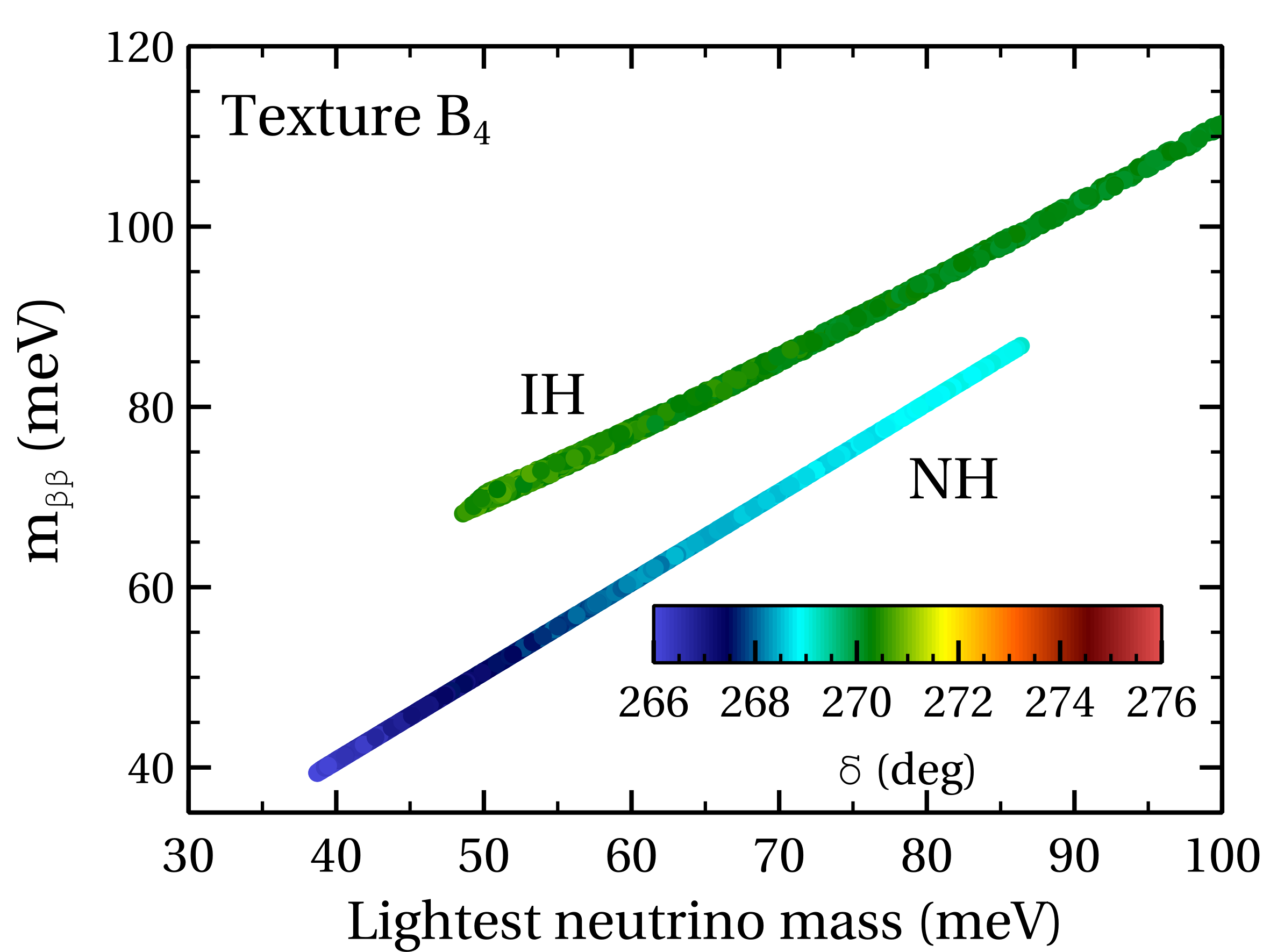}}\\
\subfigure[Texture $C$]{
\includegraphics[height=5cm,width=7.5cm]{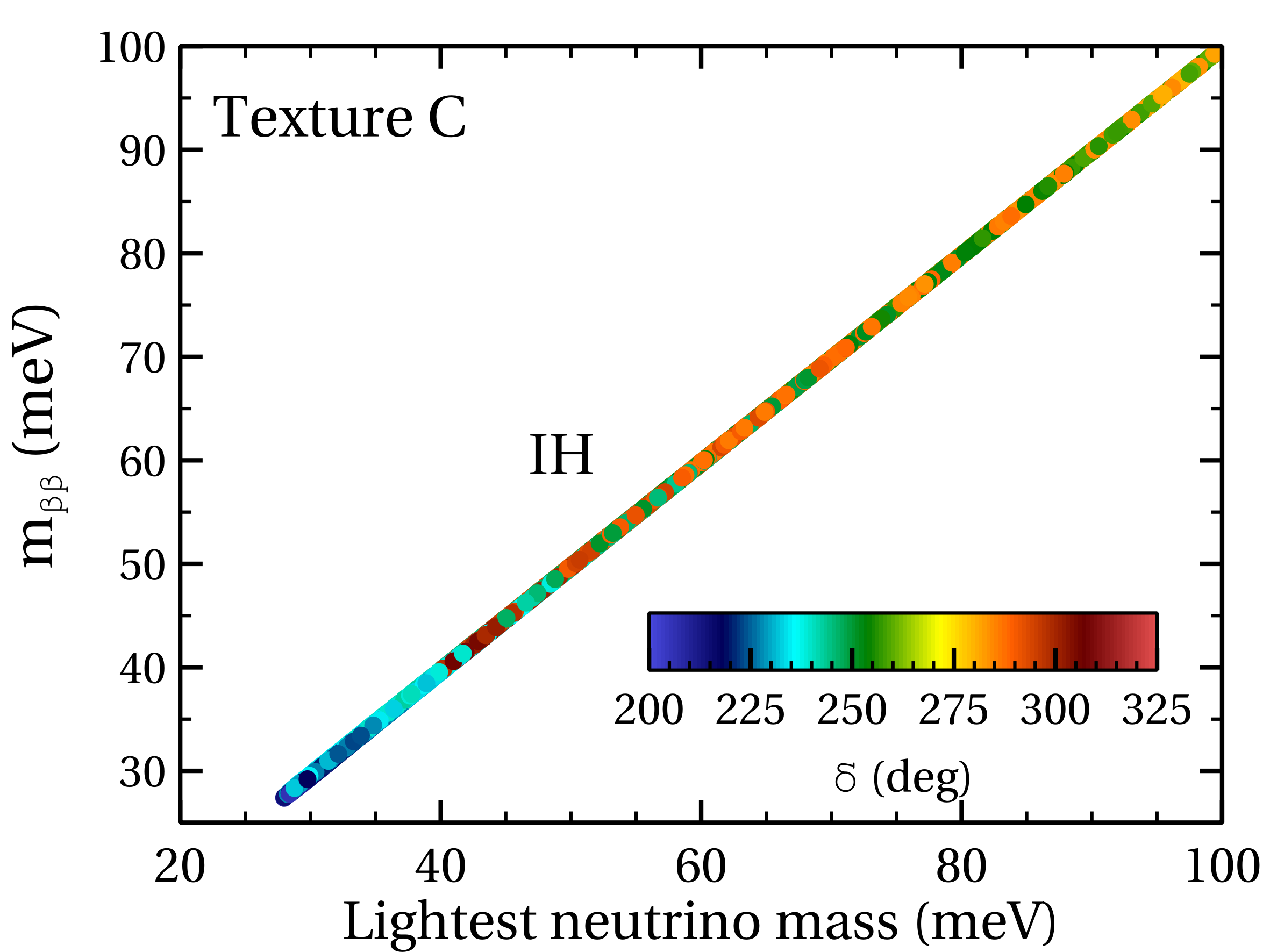}}
\caption{$m_{\beta\beta}-m_1(m_3)$ plane for all allowed two-zero textures.}
\label{fig:mbb-m3}
\end{figure}
 
Next, we have checked the possibility of probing the present scenario
at the ongoing experiments. One of the possibilities is the observation
of neutrinoless double beta decay. If the SM neutrinos are Majorana
fermion, a hypothetical radioactive process called
neutrinoless double beta decay ($^AX_Z \rightarrow ^AY_{Z+2} + 2e^{-}$)
can take place, where a parent nucleus ($X$) decays into a daughter
nucleus ($Y$) and two electrons without the emission of any neutrino.
The most important quantity that determines the double beta decay rate of the parent
nucleus is the effective Majorana mass, which is
defined as $m_{\beta\beta}  = |\sum_i V^2_{e i}\, m_i|$, where
$V = U_{\rm PMNS} \times P$ as defined in Eq.\,\eqref{eq:mdia}, and $m_i$ is the
mass eigenvalue of the $i$th neutrino. We have computed the
quantity $m_{\beta\beta}$ for all the seven possible two-zero textures
(both for normal as well as inverted mass hierarchies) and have plotted $m_{\beta\beta}$
against the lightest neutrino mass ($m_1(m_3)$ for normal(inverted) hierarchy)
in Fig.\,\,\ref{fig:mbb-m3}. The colour bar in each plot represents
the possible range of the Dirac CP phase ($\delta$) in degree for that
particular pattern of two-zero texture. From the Figs.\,\,\ref{fig:mbb-m3}(a)
and \ref{fig:mbb-m3}(b), it is seen that $m_{\beta\beta}$ is vanishingly
small for the pattern $A_1$ and $A_2$. This is primarily due to the fact that
the (1,1) element of $A_1$ and $A_2$ are zero.

For the $B$ patterns ($B_1$, $B_2$, $B_3$ and $B_4$), the effective Majorana mass
lies in the range $60\,\,{\rm meV} \lesssim m_{\beta\beta} \lesssim 100$ meV for
the inverted mass hierarchical scenario whereas for the normal mass ordering,
the corresponding range is $40\,\,{\rm meV} \lesssim m_{\beta\beta} \lesssim 90$ meV.
Finally, for the $C$ pattern, only the inverted mass ordering is experimentally allowed
and in this case we have $m_{\beta\beta}\gtrsim 25$ meV, much lower than the $B$-patterns.
Moreover, as seen from the Fig.\,\,\ref{fig:mbb-m3}, the effective Majorana mass
$m_{\beta\beta} \simeq m_3$ for the C texture. This has been shown earlier in \cite{Grimus:2004az}.
All the textures except $A_1$ and $A_2$ show a definite behaviour of $m_{\beta\beta}$
with respect to the mass of the lightest neutrino where $m_{\beta\beta}$ decreases sharply
and linearly with $m_1$ or $m_3$.
There are a few ongoing experiments which have been looking for neutrinoless double
beta decay and KamLAND-Zen \cite{KamLAND-Zen:2024eml} is one such experiment,
which uses $^{136}$Xe nucleus and has reported the most stringent lower
bound on the half-life of neutrinoless double beta decay from non-observation,
i.e. $T_{1/2} > 3.8\times10^{26}$ yr at 90\% CL \cite{KamLAND-Zen:2024eml}.
Translating the above bound on the effective Majorana mass parameter, it gives
$m_{\beta\beta} < 28-122$ meV \cite{KamLAND-Zen:2024eml}. The range in
$m_{\beta\beta}$ is due to nuclear matrix element calculations using various
models like shell model \cite{Menendez:2017fdf, Coraggio:2020hwx}, energy-density
functional theory \cite{Rodriguez:2010mn}, quasi-particle random
phase approximation \cite{Fang:2018tui}, interacting boson
model \cite{Barea:2015kwa, Deppisch:2020ztt} etc. Therefore, the double beta
decay rates predicted by different patterns of two-zero texture are
not only within the ballpark of current experimental sensitivity,
some part of the parameter space is already probed by KamLAND-Zen
depending upon the uncertainty in nuclear matrix element.
The remaining part will be tested in a very near future.  
\begin{figure}
\subfigure[Texture $A_1$]{
\includegraphics[height=5cm,width=7.5cm]{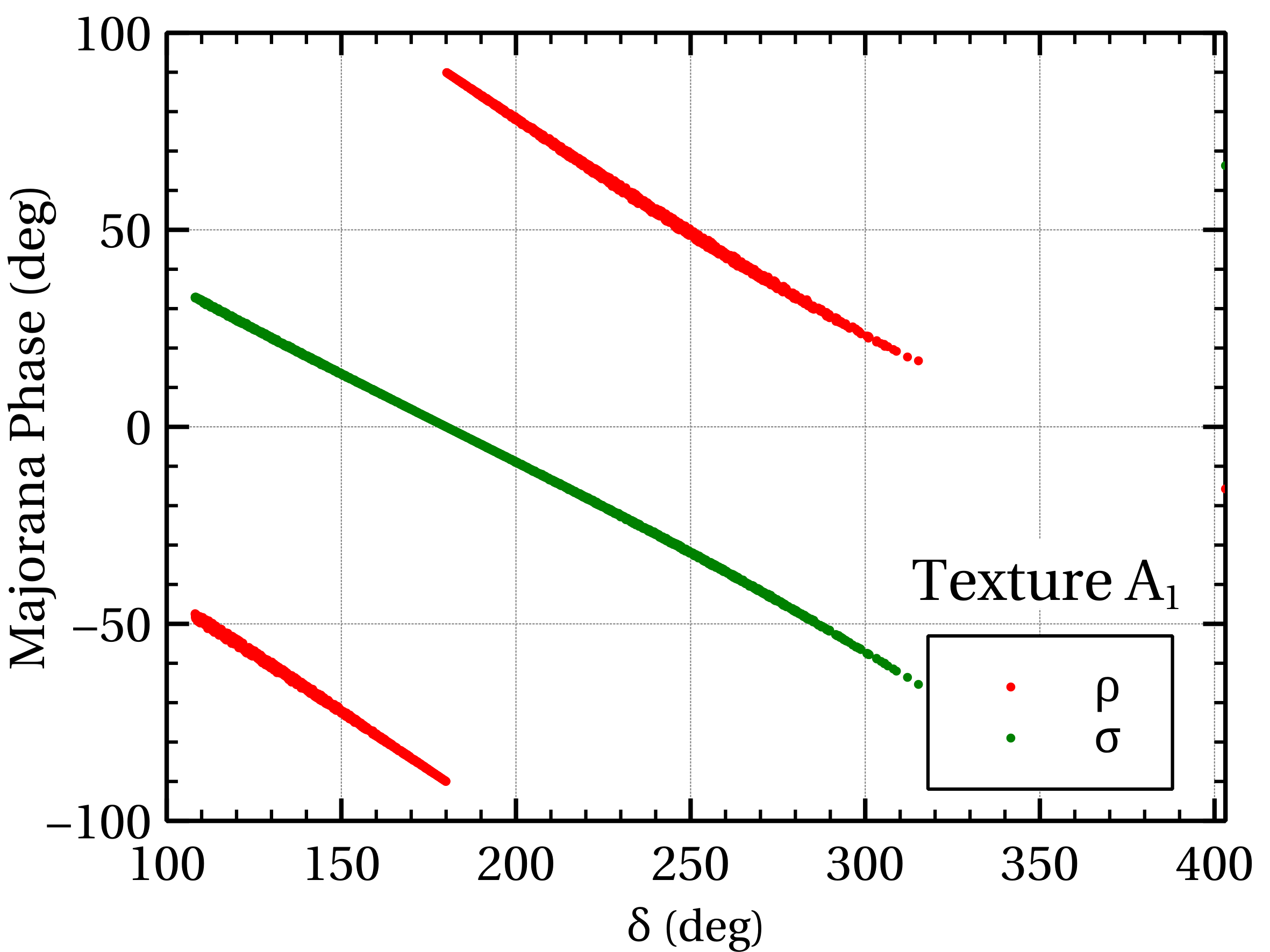}}
\subfigure[Texture $A_2$]{
\includegraphics[height=5cm,width=7.5cm]{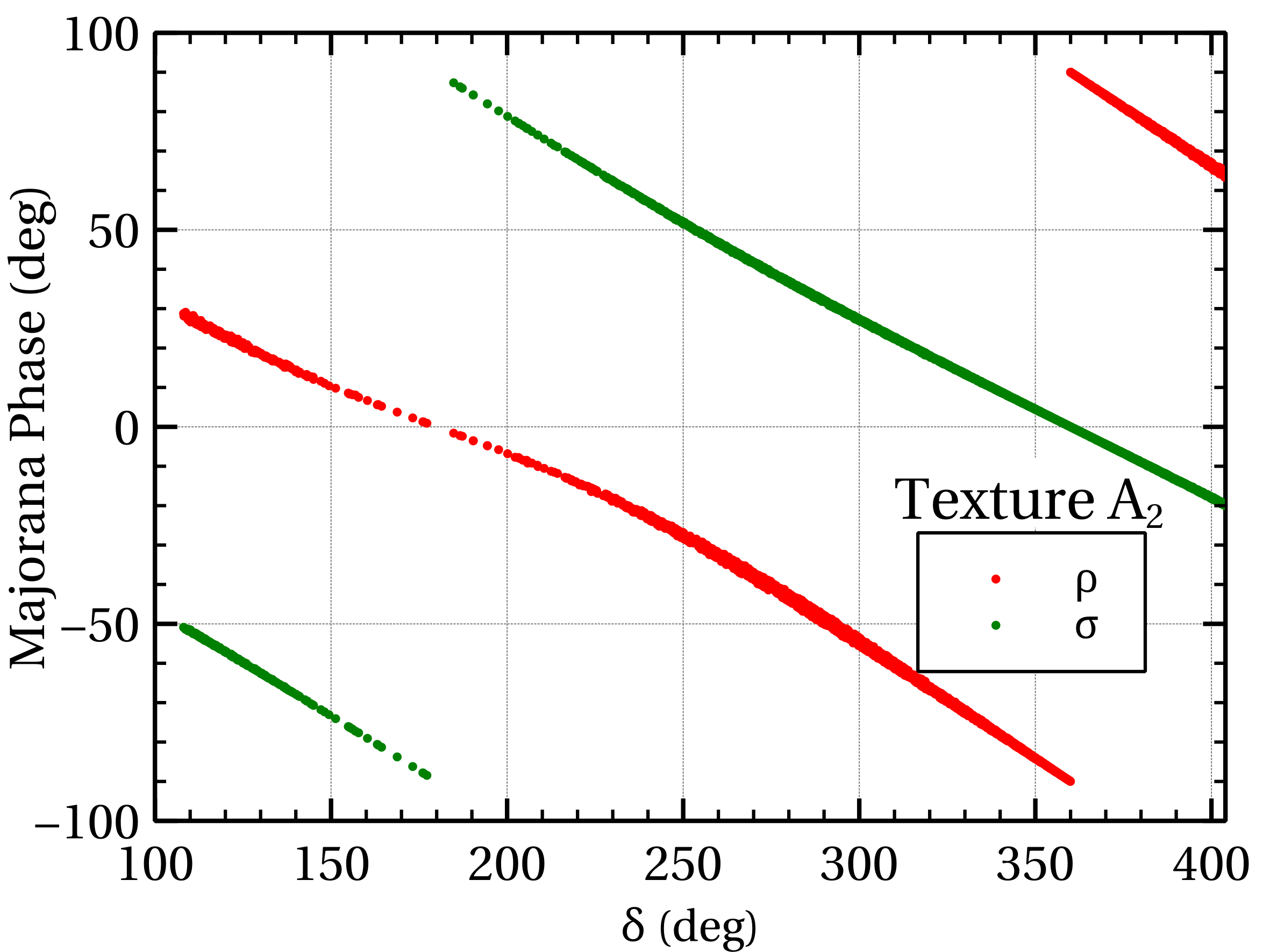}}\\
\subfigure[Texture $B_1$]{
\includegraphics[height=5cm,width=7.5cm]{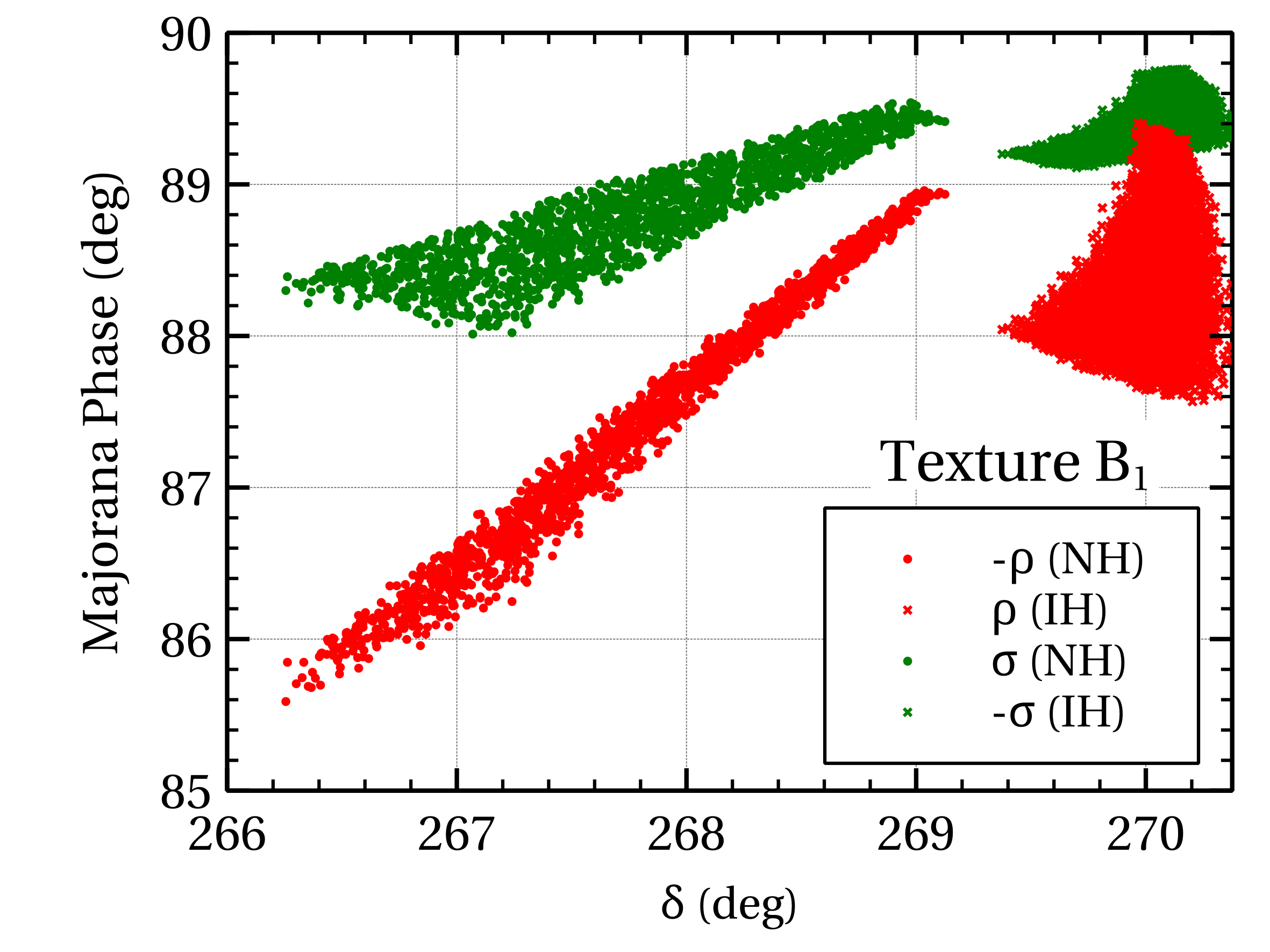}}
\subfigure[Texture $B_2$]{
\includegraphics[height=5cm,width=7.5cm]{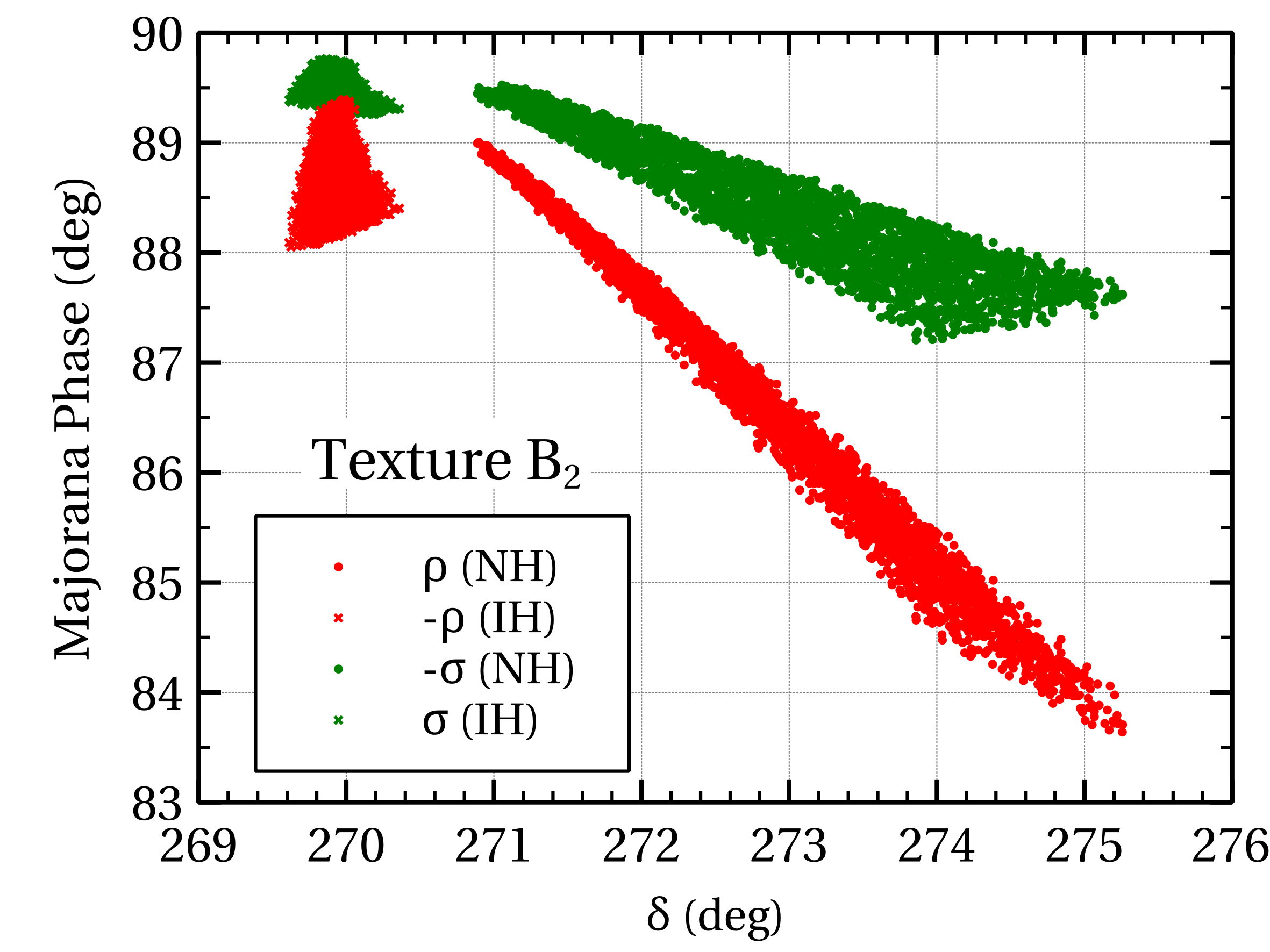}}\\
\subfigure[Texture $B_3$]{
\includegraphics[height=5cm,width=7.5cm]{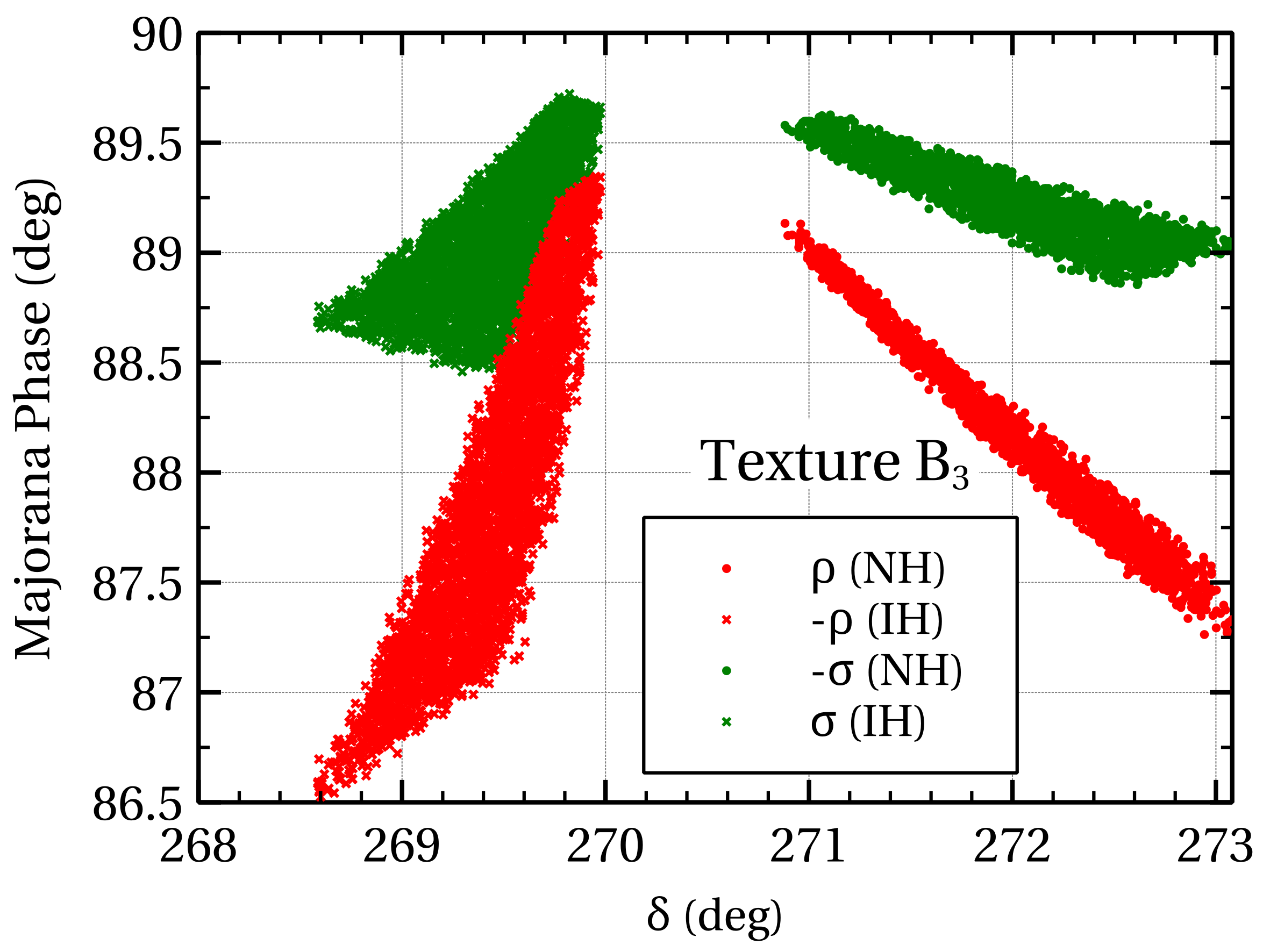}}
\subfigure[Texture $B_4$]{
\includegraphics[height=5cm,width=7.5cm]{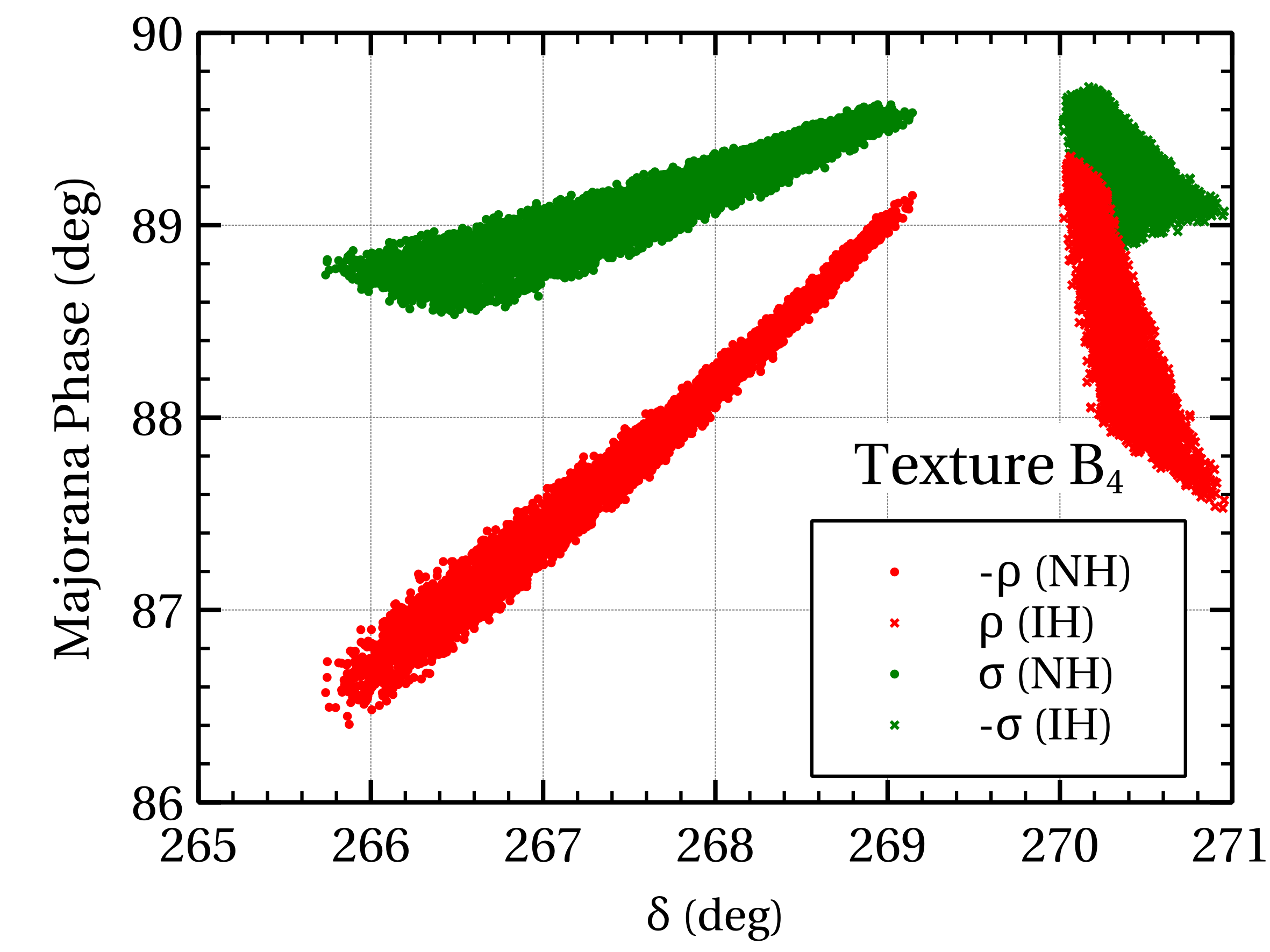}}\\
\subfigure[Texture $C$]{
\includegraphics[height=5cm,width=7.5cm]{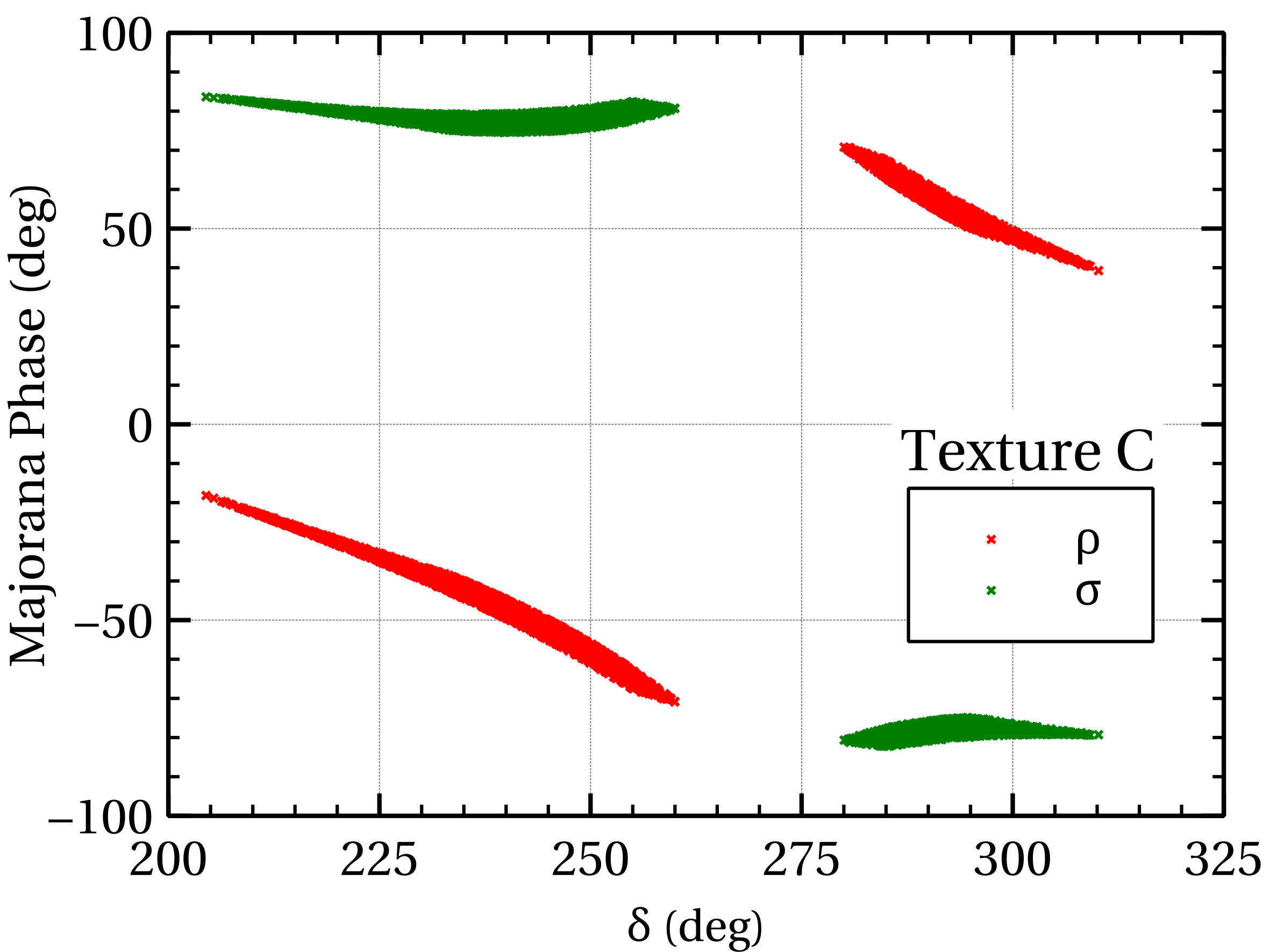}}
\caption{Predictions of the Majorana phases for all the seven allowed two-zero textures.}
\label{fig:rho-sigma-delta}
\end{figure}

{Finally, we have calculated the Majorana phases,
a unique features of Majorana neutrinos, for all seven viable two-zero textures using
Eq.\,\,\eqref{eq:majo_phases}. Our predictions are presented in Fig.\,\,\ref{fig:rho-sigma-delta},
where, for each texture, the Majorana phases ($\rho$ and $\sigma$) are shown as
functions of the allowed values of the Dirac CP phase $\delta$. In these plots,
the Majorana phase $\rho$ is represented by red points, while the green points
correspond to the other Majorana phase $\sigma$. 
Although the phases $\rho$ and $\sigma$ are computed within the interval 
$[-\pi,\pi]$, for better visualization, we have plotted 
$-\rho$ and $-\sigma$ for certain mass orderings of the B-textures.}
\section{Implementation of two-zero texture in Type-II seesaw}
\label{sec:typeII}
In this section, we will demonstrate how to implement the two-zero texture
pattern naturally in a popular neutrino mass model known as
the Type-II seesaw model \cite{Arhrib:2011uy}. The minimal Type-II
seesaw model is nothing but an extension of scalar sector
of the SM by a triplet scalar ($\Delta$) which involves in
a new Yukawa coupling with the lepton doublets. As a result,
Majorana mass is generated for the left-handed neutrinos
when $\Delta$ gets a tiny induced VEV. The Yukawa term
in Type-II seesaw model is given by
\begin{eqnarray}
\mathcal{L} \supset - i{\mathcal{Y}_{\Delta}}_{\alpha\beta}\,\ell^{T}_{\alpha} 
\mathcal{C} \sigma_2 \Delta \ell_{\beta}  + {\rm hc}\,,
\label{eq:yukawa}
\end{eqnarray}
where $\ell_{\alpha}$ is the lepton doublet of flavour $\alpha$ ($\alpha=e,\,\mu,\,\tau$),
$\mathcal{C}$ is the charge conjugation matrix and ${\mathcal{Y}_{\Delta}}_{\alpha\beta}$
is the $\alpha\beta$ element of the Yukawa coupling matrix $\mathcal{Y}_{\Delta}$, respectively. 
The neutrino mass matrix results from the above Yukawa interaction term contains
all six independent elements (complex symmetric matrix) as there is
no underlying flavour symmetry to forbid some of the Yukawa couplings.
A detailed analysis on the neutrino mass matrix in minimal Type-II
seesaw model can be found in \cite{Biswas:2017dxt}. The collider
signatures have been studied e.g. in detail~\cite{Chun:2003ej, FileviezPerez:2008jbu}. 

A definite pattern in the neutrino mass matrix is possible if
there lies a symmetry among the different neutrino flavours. This
will reduce the number of independent parameters in the neutrino mass matrix,
which otherwise are six complex numbers\footnote{Out of these six phases,
one can remove three by using the freedom of phase rotation for
the left-handed lepton doublets and the corresponding right-handed
singlets.} (or twelve real numbers) when all
the elements in $m_{\nu}$ are nonzero. A flavour symmetry in the
lepton sector is very interesting for phenomenological reason and
if it is ${\rm U(1)}_{L_{\mu}-L_{\tau}}$ symmetry then it has strong
theoretical motivation also. The ${\rm U(1)}_{L_{\mu}-L_{\tau}}$
gauge symmetry is an anomaly free extension of the SM where
the anomaly is canceled between the second and third generations
of leptons without requiring any new degrees of freedom. 
\begin{table}[h]
\begin{center}
\begin{tabular}{|c|c|c|c|c|c|}
\hline
\hline
Texture Type & $L_{\mu}-L_{\tau}$&$L_{\mu}-L_{\tau}$&$L_{\mu}-L_{\tau}$& Allowed by & Allowed by \\
& charge of & charge of & charge of & NH & IH \\
& $\Delta_1$ & $\Delta_2$ & $\Delta_3$ & & \\
\hline
\hline
$A_1$ & \multicolumn{3}{c|}{Not possible to generate} & \cmark  & \xmark \\
& \multicolumn{3}{c|}{(\small more than two independent zeros appear)} & & \\ 
\hline 
$A_2$ & \multicolumn{3}{c|}{Not possible to generate}   &    \cmark & \xmark \\
& \multicolumn{3}{c|}{(\small more than two independent zeros appear)} & & \\ 
\hline 
$B_1$ &      0       &    -1        &    2        & \cmark & \cmark \\
\hline
$B_2$ &      0       &     1       &    -2        & \cmark & \cmark \\
\hline 
$B_3$ &      0       &     1      &      2       & \cmark & \cmark \\
\hline 
$B_4$ &  0           & -1           &  -2          & \cmark & \cmark \\
\hline 
$C$ &   0          &   1         &   -1         & \xmark & \cmark \\
\hline
\hline 
\end{tabular}
\caption{The charge of triplets under ${\rm U}(1)_{L_{\mu}-L_{\tau}}$ symmetry.}
\label{tab:lmu-ltau_charges}
\end{center}
\end{table}  
 
However, within the minimal Type-II seesaw framework, it is not so straight
forward to impose $L_{\mu}-L_{\tau}$ symmetry directly since
that will jeopardize the structure of neutrino mass matrix completely.
For example, if we do not assign any $L_{\mu}-L_{\tau}$ charge
to $\Delta$, the first generation will be decoupled from the other
two and the mixing angle $\theta_{23} = 45^\circ$. Assigning
any nonzero charge to $\Delta$ will not help much; rather, it will
make the situation even worse. One of the elegant ways to achieve
the oscillation parameters in the appropriate range (as given
in Table \ref{tab:oscillation_data}) along with a definite flavour  
structure in $m_{\nu}$ requires more triplets in the theory
(see Table\,\ref{tab:no_of_triplet}). As we
mentioned earlier, presence of an underlying symmetry among
the neutrino flavours will reduce the number of independent
parameters in $m_{\nu}$. The current neutrino oscillation data
can allow maximum of two independent elements to be zero. The
minimal number of parameters in $m_{\nu}$ (i.e. the two-zero structure)
can be naturally achieved when we have at least three triplets ($\Delta_1$,
$\Delta_2$ and $\Delta_3$) in the scalar sector, along with
our Higgs doublet responsible for the electroweak symmetry
breaking. In this three triplet framework, we can easily obtain
different patterns of the two-zero textures like
$B_1$, $B_2$, $B_3$, $B_4$ and $C$, and in each case
we need different $L_{\mu}-L_{\tau}$ charges for the three
triplets. For example, the pattern $C$ can be achieved if
we assign $0$, $1$ and $-1$ charges to $\Delta_1$, $\Delta_2$
and $\Delta_3$, respectively. For the other patterns, the necessary
$L_{\mu}-L_{\tau}$ charges of the triplets are listed in
Table\,\ref{tab:lmu-ltau_charges}. Moreover, in this scenario, it
is not possible to obtain the $A$ texture ($A_1$ and $A_2$)
in the neutrino mass matrix since the pattern of $L_{\mu}-L_{\tau}$
charges enforces $(m_{\nu})_{23}$ and $(m_{\nu})_{32}$ elements to be zero
along with the $(m_{\nu})_{11}$ element. Therefore, more than two
independent zeros will appear in $m_{\nu}$ since $1\times 2$ and $1\times 3$
elements are also zero in $A_1$ and $A_2$, respectively. 
In the last two columns of Table\,\ref{tab:lmu-ltau_charges},
we have shown whether a particular pattern favours either
the normal mass ordering or the inverted mass ordering or both.        
\section{Constraints from experiments}
\label{sec:constaints}
There are several constraints on the Yukawa
couplings (Eq.\,\eqref{eq:yukawa}) from lepton
flavour violating three body decays like
$\tau \rightarrow \bar{\ell_i} \ell_j \ell_k$
and $\mu \rightarrow \bar{e} e e$. In the present
scenario, since we have three triplets, all these
processes except $\tau \rightarrow \bar{\mu} e e$
depend on how large are the mixing angles among different triplets. Since
a detailed analysis on the scalar sector Lagrangian is beyond the
scope of this article and is presented in a separate paper \cite{Biswas2025xp}, we
have not considered those processes that are mixing suppressed. The 
decay channel $\tau \rightarrow \bar{\mu} e e$ is independent of
triplet mixing and it occurs through the off-shell production
of doubly charged scalar of the triplet $\Delta_1$ (having zero
$L_{\mu}-L_{\tau}$ charge) and its subsequent decay into a pair of electrons.
The Branching ratio of this decay channel
is given by \cite{Kakizaki:2003jk, Antusch:2018svb}
\begin{eqnarray}
{\rm Br}(\tau \rightarrow \bar{\mu} e e) =
\dfrac{1}{64 G^2_{F} M_1^4}
\left|{\mathcal{Y}_{\Delta}}_{32}\,{\mathcal{Y}_{\Delta}}_{11}\right|^2\,,
\label{eq:tau2mubaremep}
\end{eqnarray}
where $M_1$ is the mass of the doubly charged scalar
in triplet $\Delta_1$ and $G_{F}$ is the Fermi constant.
The current bound is ${\rm Br}(\tau \rightarrow \bar{\mu} e e)<1.5\times 10^{-8}$ \cite{Hayasaka:2010np}.
This results in an upper bound on the product of two
Yukawa couplings from Eq.\,\eqref{eq:tau2mubaremep} as 
\begin{eqnarray}
\left|{\mathcal{Y}_{\Delta}}_{32}\,{\mathcal{Y}_{\Delta}}_{11}\right|
< 0.0114 \left(\dfrac{M_1}{{\rm TeV}}
\right)^2 \,.
\end{eqnarray}
There are also lepton flavour violating two body
decays such as $\mu \rightarrow e \gamma$,
$\tau \rightarrow \mu \gamma$ and $\tau \rightarrow e \gamma$.
In presence of singly and doubly charged scalars, these rare
decays are possible at one loop level where these charged scalars
and the corresponding leptons\footnote{for singly(doubly) charged scalar the
corresponding lepton is neutrino(charged lepton).} are within the loop.
However, as we mentioned before, this requires mixing among the triplets
since all the leptons are not coupled to a single triplet via the Yukawa
coupling (Eq.\,\eqref{eq:yukawa}) due to specific $L_{\mu}-L_{\tau}$
charge assignment (Table\,\ref{tab:lmu-ltau_charges}). For $\mu \rightarrow e \gamma$,
for the case of $B_1$ texture, we need mixing between $\Delta_1$ and $\Delta_2$
while for $C$ texture mixing between $\Delta_1$ and $\Delta_2$
or/and $\Delta_1$ and $\Delta_3$ is required. Therefore, the
constraints on the Yukawa couplings obtained from these rare two body
decays are usually suppressed by relevant mixing angles and it
requires a detailed analysis of the multi-field scalar potential,
which we have done in a separate work \cite{Biswas2025xp}.   
 
We would like to note that in minimal Type-II seesaw model, the
triplet VEV is an induced VEV which is generated when the Higgs
doublet gets a VEV and the elctroweak symmetry is broken spontaneously.
This happens due to a trilinear term between the two Higgs doublets
and the triplet like $\mu_T H^T i\sigma_2 \Delta^{\dagger} H$ which is
allowed by SU(2)$_{L} \otimes {\rm U(1})_Y$ symmetry. However,
in the present scenario, since two triplets have nonzero $L_{\mu}-L_{\tau}$
charge (see Table\,\ref{tab:lmu-ltau_charges}), the similar trilinear term
is possible for the triplet $\Delta_1$ only. Therefore, $\Delta_1$
will get an induced VEV after the electroweak symmetry breaking. However,
no VEV will be generated for the other two triplets ($\Delta_2$ and
$\Delta_3$) as it is protected by the $L_{\mu}-L_{\tau}$ symmetry.
The only way to achieve nonzero VEVs is to break the $L_{\mu}-L_{\tau}$
symmetry and one of the simple and elegant ways to do this
is by adding an SU(2)$_{L}$ singlet scalar charged under
the $L_{\mu}-L_{\tau}$ symmetry. Therefore, with the help of
singlet scalar $\phi$, one can write the following
interaction terms $\lambda_T H^T i\sigma_2 \Delta^\dagger_2 H \phi$
and $\lambda_{m} {\rm Tr}\left({\Delta^{\dagger}_1 \Delta_3}\right) (\phi^{\dagger})^2$
for the texture $B_1$, where the $L_{\mu}-L_{\tau}$ charge of $\phi$ is
+1. When $\phi$ gets a VEV, the first term will
generate trilinear term for $\Delta_2$, while
the second term will generate a nonzero mixing between $\Delta_1$
and $\Delta_3$, which eventually generates an effective trilinear
term for the triplet $\Delta_3$ also. A similar procedure
can be followed for the other $B$ textures ($B_2$, $B_3$ and $B_4$),
where the only difference will be the $L_{\mu}-L_{\tau}$ charge of
$\phi$. For $C$ texture, the $L_{\mu}-L_{\tau}$ charge of triplets
allows us to write the quartic interaction terms like
$\lambda_{2T} H^T i\sigma_2 \Delta_2^\dagger H \phi^{\dagger}$
and $\lambda_{3T} H^T i\sigma_2 \Delta^\dagger_3 H \phi$ which later generate
trilinear terms for both the triplets $\Delta_2$ and $\Delta_3$ simultaneously
after U(1)$_{L_{\mu}-L_{\tau}}$ breaking.       
\section{Conclusion}
\label{sec:conclusion}
In this work, we have reanalyzed the two-zero texture of the neutrino mass
matrix. There are fifteen different types of two-zero textures, out
of which only seven are allowed by the neutrino oscillation data. These textures
are labeled as $A_1, A_2, B_1, B_2, B_3, B_4$ and $C$ respectively.
Moreover, all these seven textures are not independent, as some of them
are related by a permutation symmetry like $A_2 = P^T_{23} A_1 P_{23}$,
where $P_{23}$ is an orthogonal matrix. As a result, there are only
four independent two-zero patterns namely, $A_1, B_1, B_3$ and $C$.
We have calculated observables related to the neutrino oscillation
such as $\Delta{m}^2_{21}$, $\Delta{m}^2_{3\ell}$, $\theta_{12}$,
$\theta_{23}$, $\theta_{13}$ and $\delta$ for all the seven
allowed two-zero textures numerically. Our prediction on
the Dirac CP phase ($\delta$) and the absolute mass scale of
neutrinos are demonstrated in $\delta-m_3$ 
plane (Fig.\,\ref{fig:delta-m3}) with the sum of neutrino masses is shown
using a colour bar. We have observed that the Dirac CP phase ($\delta$) is
confined to very narrow ranges, i.e. $265^\circ \lesssim \delta \lesssim 275^\circ$ for $B$
textures and $205^\circ \lesssim \delta \lesssim 260^\circ$
and $280^\circ \lesssim \delta \lesssim 310^\circ$ for $C$ texture respectively, which
could be a smoking gun signature for these two-zero textures if
we are able to measure $\delta$ precisely in the upcoming oscillation
experiments like DUNE \cite{DUNE:2015lol} and
Hyper-Kamiokande \cite{Hyper-Kamiokande:2018ofw}. However,
for texture $A_2$, almost the entire range of $\delta$
(see Table\,\ref{tab:oscillation_data}) is allowed while
for $A_1$, the CP phase lies between $\sim 120^\circ$ and
$\sim 330^\circ$. 

We have found that the $A$ textures prefer
the normal mass ordering, while the inverted mass ordering is preferred
by $C$ texture. On the other hand, both normal and inverted
mass hierarchies could be a possibility for the $B$ textures. Moreover,
the detection prospects of two-zero textures have been explored
by computing the effective Majorana mass ($m_{\beta\beta}$) for all the
textures in Fig.\,\,\ref{fig:mbb-m3}, which is a key parameter in experiments searching for  
the neutrinoless double beta decay. We have found that $m_{\beta\beta}$
for the $B$ and $C$ textures are not only within the experimental sensitivity of
KamLAND-Zen but also some of the parameter space in $m_{\beta\beta}-m_i$
(where $i=1(3)$ for NH(IH)) plane is already excluded. 
Future results on the lifetime of neutrinoless double
beta decay will easily probe the entire parameter space of
two-zero textures very soon and may thereby shed light on
the nature of neutrinos. Moreover, we have computed
the Majorana phases ($\rho$, $\sigma$) for all seven textures
and our predictions are illustrated in Fig.\,\,\ref{fig:rho-sigma-delta}.   

In the remaining part of the work, we have concentrated on
implementation of the two-zero pattern in a realistic
neutrino mass model. The neutrino mass matrix in minimal Type-II
seesaw scenario has all the elements, and therefore does not have
any flavour structure. This results in too many unwanted parameters
in $m_{\nu}$. Adding a flavour symmetry like U(1)$_{L_{\mu}-L_{\tau}}$
which has other theoretical motivations in particle physics is very interesting.
However, this requires additional scalar triplets as the minimal Type-II
model with ${L_{\mu}-L_{\tau}}$ symmetry is unable to reproduce the appropriate
structure of the neutrino mass matrix. A realistic neutrino mass matrix can
be obtained if we have at least three triplets with
different ${L_{\mu}-L_{\tau}}$ charges (see Table\,\ref{tab:no_of_triplet}).
Most interestingly, this scenario automatically leads to the
two-zero texture pattern in $m_{\nu}$, which we have studied
extensively in the first part. We have found that
all the allowed two-zero patterns except $A$ textures
can be generated in this scenario. However, for the $A$
textures more than two independent zeros appear
in $m_{\nu}$, which is disallowed by the oscillation
experiments. Moreover, the U(1)$_{L_{\mu}-L_{\tau}}$
symmetry forbids some of the trilinear couplings between
triplet scalar and the SM Higgs doublet, which is
crucial for neutrino mass generation. We have discussed a simple
and elegant way to generate the trilinear couplings. Finally, we have
discussed various constraints coming from lepton flavour violating 
decays on the new Yukawa couplings and have found that the constrains
are predominantly coming from three body decay $\tau \rightarrow \bar{\mu} ee$
since all the other lepton flavour violating processes heavily depend
on the mixing among the triplet scalars, which requires an extensive
study of the full scalar potential involving three triplets
along with our SM Higgs doublet. We have explored this in
a separate paper \cite{Biswas2025xp}.        
\section{Acknowledgements} 
This work is supported by the National Research Foundation
of Korea (NRF) grant funded by the Korea government (MSIT)
RS-2023-00283129 and RS-2024-00340153. AB would like to thank
Mainak Chakraborty for some useful discussions. 
The research of SJ was supported by an appointment to the YST Program
at the APCTP through the Science and Technology Promotion Fund and Lottery
Fund of the Korean Government. This was also supported by the Korean
Local Governments - Gyeongsangbuk-do Province and Pohang city (S.J.).
\bibliographystyle{JHEP}
\providecommand{\href}[2]{#2}\begingroup\raggedright\endgroup

\end{document}